\documentclass[11pt]{article}

\usepackage{natbib}
\usepackage{amsmath}
\usepackage{graphicx}
\usepackage[colorinlistoftodos]{todonotes}
\usepackage[colorlinks=true, allcolors=black]{hyperref}
\usepackage{amsmath,amssymb,amsthm,bm,bbm,fullpage}
\usepackage{mathrsfs}
\usepackage{colortbl}
\usepackage[all]{xy}
\usepackage{graphics}
\usepackage{multirow}
\usepackage[affil-it]{authblk}
\usepackage{lineno}
\renewcommand{\citep}[1]{\citeauthor{#1}~\citepnum{#1}}

\usepackage{setspace}
\usepackage[font=footnotesize]{caption}

\usepackage{xr-hyper}
\usepackage{hyperref}
\theoremstyle{remark} 
	
\providecommand{\keywords}[1]
{
  \small	
  \textbf{\textit{Keywords---}} #1
}

\newenvironment{sciabstract}{%
\begin{quote}\bf}
{\end{quote}}

\usepackage{xspace}
\newcommand{\SI}{Supplementary Material\xspace}

\title{Community Notes are Vulnerable to Rater Bias and Manipulation}

\author[1,2,*]{Bao Tran Truong}
\author[2]{Siqi Wu}
\author[1,2]{Alessandro Flammini}
\author[1,2]{Filippo Menczer}
\author[1,2,*]{Alexander J. Stewart}

\affil[1]{Observatory on Social Media, Indiana University}
\affil[2]{Luddy School of Informatics, Computing, and Engineering, Indiana University}
\affil[*]{baotruon@iu.edu,stewalex@iu.edu}

\begin{document}

\date{}
\maketitle

\begin{singlespace}
\begin{sciabstract}
Social media platforms increasingly rely on crowdsourced moderation systems like Community Notes to combat misinformation at scale. However, these systems face challenges from rater bias and potential manipulation, which may undermine their effectiveness. Here we systematically evaluate the Community Notes algorithm using simulated data that models realistic rater and note behaviors, quantifying error rates in publishing helpful versus unhelpful notes. We find that the algorithm suppresses a substantial fraction of genuinely helpful notes and is highly sensitive to rater biases, including polarization and in-group preferences. Moreover, a small minority (5--20\%) of bad raters can strategically suppress targeted helpful notes, effectively censoring reliable information. These findings suggest that while community-driven moderation may offer scalability, its vulnerability to bias and manipulation raises concerns about reliability and trustworthiness, highlighting the need for improved mechanisms to safeguard the integrity of crowdsourced fact-checking.
\end{sciabstract}
\end{singlespace}

\keywords{Community Notes, Social Media Moderation, Crowdsourcing, Rater Bias, Polarization, Algorithm Manipulation}

\clearpage

Over the past decade, online platforms have increasingly struggled with how to moderate misinformation at scale. Traditional approaches, such as professional fact-checking and content labeling, have been shown to effectively reduce the spread of false information and increase the accuracy of user beliefs \cite{Hoes24,fazio2024mega}. Despite this demonstrated effectiveness, many major platforms have recently moved away from expert-driven methods in favor of ``community-based'' moderation systems.

The most prominent demonstration of this shift is Community Notes, a crowdsourced approach to fact-checking introduced by Twitter (now X) as Birdwatch, and later expanded under its current name \cite{cn-guide}. Community Notes allows platform users to collaboratively identify and contextualize potentially misleading content. Its open-source algorithm for selecting and publishing ``helpful'' notes is designed to promote consensus across ideological lines \cite{wojcik2022birdwatch}, representing an effort to balance transparency, participation, and perceived fairness.

Following X, other platforms, including Facebook and TikTok, have begun to replace traditional fact-checking programs with versions of Community Notes, at least in the U.S. context. Both have adopted the same core algorithm as X, often citing its purported ability to reduce bias and increase user trust. While empirical support for the claim of reduced bias remains limited, some studies do suggest that community-driven annotations can increase user confidence in moderation outcomes \cite{drolsbach2024community}. As of July 2025, TikTok reported more than 80,000 contributors to its Footnotes program \cite{TikTok-Footnotes}, while Meta announced that its Community Notes initiative had generated around 15,000 notes from 70,000 users since its March 2025 launch \cite{Meta-Threads}.

This enthusiasm reflects broader optimism about the scalability and effectiveness of crowdsourced fact-checking. 
Early research finds that community fact-checkers can achieve accuracy comparable to professional fact-checkers \cite{allen2021scaling,BARBERA2024103792}, and that the approach may be more scalable and cost-effective than expert-based models \cite{Martel24crowdscale,Saeed22}. 
Exposure to community-authored corrections seems less likely to provoke disengagement than individual misinformation tagging \cite{kim2025differential}, even though layperson volunteers are perceived as less legitimate than experts \cite{Resnick2025legitimacy}. 
Some studies suggest that when shown to users, Community Notes reduce engagement with and diffusion of false or misleading content \cite{drolsbach_diffusion_2023,kankham_community_2024,slaughter2025community}. 
There is weak consensus on this finding, however, especially when considering engagement with misinformation in the early stage of diffusion \cite{chuai2024did} and misinformation originating from influential users \cite{prollochs2022community}.  

The performance of community moderation in real-world conditions is uneven. Crowdsourcing introduces challenges of coverage, timeliness, and reliability. Volunteer contributors may self-select content that reflects their interests or ideological motivations, and may draw on inconsistent sources when writing notes \cite{allen2022birds,Saeed22}. These limitations, combined with platform-level visibility constraints, raise concerns about whether community moderation can provide an effective large-scale substitute for traditional fact-checking \cite{augenstein2025communitymoderationnewepistemology}.

Visibility is a central challenge. According to one study~\cite{ccdh}, 74\% of misleading posts on X that had accurate notes were never shown to users. 
Investigations have found that most users never encounter published notes at all \cite{Mashable}. 
When notes are shown, they receive much lower exposure than the corresponding posts \cite{ccdh,Mashable,slaughter2025community}.

An explanation for this gap in visibility and impact is the delay with which notes appear when they are published. 
Between January 2021 and January 2025, only 13.55\% of posts with at least one proposed note on X received a ``helpful'' note, while 87.7\% remained in the ``Needs More Ratings'' category. Of all notes submitted, just 8.3\% achieved ``helpful'' status and were published, with an average delay of 26 hours \cite{mohammadi2025birdwatch} --- well past the period of peak engagement on most platforms \cite{brashier2021timing,truong2025illegaltakedowndelay}. Meta reports similar figures, with only about 6\% of notes ultimately published \cite{Meta-Threads}. 

Trends on X further suggest a decline in published notes. The share of ``helpful'' notes fell from a peak of 13.1\% in March 2024 to 8.2\% in May 2025. At the same time, contributor retention is decreasing: the proportion of authors remaining active six months after writing their first note dropped from 63\% in early 2023 to 44\% in late 2024. Because contributors whose notes are never published are less likely to continue participating, this dynamic raises concerns about the long-term sustainability of the system \cite{Alexios25}.

Beyond questions of efficacy, scholars have highlighted deeper risks in replacing expert fact-checking with crowdsourced moderation. These include the potential for manipulation, the erosion of epistemic authority, and the emergence of politically motivated behavior within crowdsourced systems \cite{augenstein2025communitymoderationnewepistemology}. In polarized online environments, where users often lack incentives for cross-ideological collaboration, community-driven moderation may reflect or even amplify partisan bias \cite{Yasseri2021Wiki}. Reports of politically motivated down-rating \cite{Wired} and evidence that contributors are often motivated by correcting ideological opponents \cite{allen2022birds} suggest that the aspiration toward neutrality may be difficult to sustain in practice.

These developments position community moderation at the center of a critical debate: whether crowdsourced intelligence can serve as a trustworthy, scalable, and robust alternative to expert-led content moderation. 
Here we contribute to this debate by analyzing the vulnerabilities of community moderation to rater bias and manipulation. 
Since neither Meta nor TikTok provides access to their data or algorithms, we use X's open-source Community Notes algorithm. 
By feeding it with simulated ratings that model reasonable assumptions about notes and raters, we can systematically and precisely quantify different kinds of errors, such as helpful notes that are not published and published notes that are not helpful. 
This analysis reveals that the Community Notes algorithm is highly sensitive to rater biases, such as the observed tendency to rate notes that are not aligned with one's political views. 
The algorithm is also vulnerable to manipulation: a small fraction of bad raters can easily suppress helpful notes so that they are never published. 

\section*{Results}

\subsection*{Experimental design}

We focus on the efficacy of the Community Notes algorithm as described in Wojcik et al. \cite{wojcik2022birdwatch} and in the release notes accompanying the open-source code \cite{cn-guide}. 
A note $n$ receives a rating $r_{un}$ from a rater $u$. 
Ratings take the value of either 1 (``HELPFUL''), 0.5 (``SOMEWHAT HELPFUL''), or 0 (``NOT HELPFUL''). 
The algorithm \emph{models} the rating as
\(
\hat{r}_{un}=\hat{\mu}+\hat{i}_u+\hat{i}_n+\hat{f}_u\hat{f}_n
\)
where $\hat{\mu}$ accounts for the overall tendency of all raters to rate notes as helpful, $\hat{i}_u$ is the tendency of rater $u$ to rate notes as helpful (their friendliness), $\hat{i}_n$ is the helpfulness of note $n$, $\hat{f}_u$ is the bias of rater $u$, and $\hat{f}_n$ is the bias of note $n$. 
Following the current implementation of the algorithm, which assumes one-dimensional bias, the $\hat{f}$ terms are scalar. 
The algorithm attributes HELPFUL ratings to a combination of three factors: (1)~the raters are lenient (high $\hat{i}_u$), (2)~the note is truly helpful (high $\hat{i}_n$), and/or (3)~the rater bias and note bias align (high $\hat{f}_u\hat{f}_n$). If a note receives many HELPFUL ratings from raters with diverse biases, the individual contributions of $\hat{f}_u\hat{f}_n$ will tend to counteract each other, because some raters will over-estimate and some will under-estimate the note's helpfulness. Then, assuming that the note is not rated by a group of particularly lenient raters, this note must be truly helpful. 

\begin{figure*}
\centering
\includegraphics[width=1.0\textwidth]{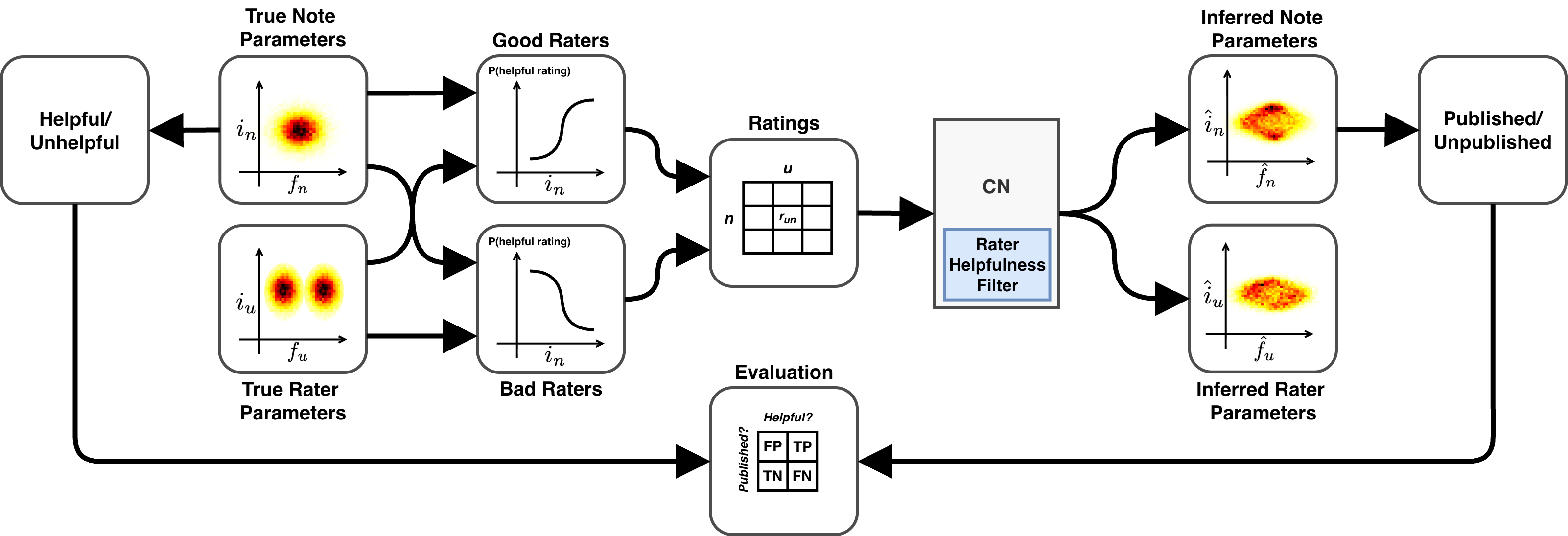}
\caption{\textbf{Experimental design.} We simulate raters and notes using the same assumptions as the Community Notes algorithm. Each note $n$ has two true parameters: helpfulness $i_n$ and bias $f_n$. Similarly, each rater $u$ is characterized by true friendliness $i_u$ and bias $f_u$. These parameters are drawn from distributions that reflect real-world features of the populations that write and rate notes, such as rater polarization and opinion diversity. 
Raters assign ratings to notes based on the helpfulness and bias of the note, as well as on their own friendliness and bias. To study the effects of manipulation, we distinguish between ``good'' and ``bad'' raters. Good raters behave as intended by the Community Notes algorithm --- they assign higher ratings to notes they perceive as helpful. In contrast, a minority of bad raters intentionally down-rate helpful notes. The Community Notes algorithm uses the simulated ratings we generate to estimate inferred parameters for notes, $\hat{i}_n$ and $\hat{f}_n$, and raters, $\hat{i}_u$ and $\hat{f}_u$. It then uses the inferred note parameters to decide whether a note should be published. In our evaluation, we compare the algorithm's publication decisions using inferred parameters to those based on the true parameters ($i_n$ and $f_n$), representing the ideal case of perfect parameter inference.}
\label{fig:0}
\end{figure*}

The Community Notes algorithm estimates the values of $\hat{\mu}$, $\hat{i}_u$, $\hat{i}_n$, $\hat{f}_u$, and $\hat{f}_n$ using matrix factorization (see \nameref{sec:methods}). 
The algorithm also has a ``helpfulness filter'' that removes raters whose ratings consistently mis-align with the status decision (see \nameref{sec:methods}). 
To study the efficacy of the algorithm, we generate simulated rating data using the same assumptions as the algorithm. 
Figure~\ref{fig:0} summarizes our experimental framework. We assume each rater $u$ is characterized by two parameters, $i_u$ and $f_u$, which describe their true propensity to rate notes as helpful (rater friendliness), and their true (one-dimensional) bias. Similarly, each note $n$ is characterized by $i_n$ and $f_n$, which describe its true helpfulness and (one-dimensional) bias. Finally a (fixed) global parameter $\mu$ determines the overall tendency of raters to rate notes as helpful.
In each simulated rating dataset, the parameters $\mu$, $i_u$, $i_n$, $f_u$, and $f_n$ are set or drawn from distributions formulated to capture some realistic scenarios. 
We assume raters employ a softmax activation function to determine how to rate a note given these parameters.
The function is tuned to accurately reproduce the mix of ratings in the empirical data (see \nameref{sec:methods}). 

We run the open-source Community Notes algorithm on simulated datasets to assess its accuracy. Each simulated dataset comprises 20,000 notes and 10,750 raters, with 1,839,726 ratings distributed to match the empirical degree distribution of the rater-note network (see \nameref{sec:methods}). Our results are robust with respect to the size of the dataset [see \SI Section~\ref{sec:si-robustness-datasize}].


Our key concern is the accuracy of the algorithm's inferences of $\hat{f}_n$ and $\hat{i}_n$, which are used to determine whether a note should be published. The basic criterion used by the Community Notes algorithm is that a note $n$ is published only if $\hat{i}_n>0.4$ and $|\hat{f}_n|<0.5$, that is, it is sufficiently helpful and not too biased. 
By comparing the inferred values $\hat{f}_n$ and $\hat{i}_n$ to the true values $f_n$ and $i_n$, we are able to determine the error rate of the algorithm for a given simulated dataset. 
We report on four types of potential errors:
the \textit{suppression rate} is the fraction of helpful notes that are not published; 
the \textit{pollution rate} is the fraction of published notes that are unhelpful;  
the \textit{infiltration rate} is the fraction of unhelpful notes that are published; 
and the \textit{waste rate} is the fraction of unpublished notes that are helpful. 
We first explore how all four of these error rates change with the underlying distribution of note and rater parameters, $f_n$, $i_n$, $f_u$ and $i_u$. We then explore how the error rates are impacted by ``bad raters,'' who deliberately mis-rate notes they perceive to be truly helpful.

\subsection*{Vulnerability to rater and note biases}

We begin by exploring the relationship between the underlying distribution of note helpfulness and bias, $i_n$ and $f_n$, rater friendliness and bias $i_u$ and $f_u$, and the accuracy of the Community Notes algorithm. Rater friendliness and note helpfulness, $i_u$ and $i_n$, are drawn independently from Gaussian distributions with means $\mu^I_j$ and standard deviation $\sigma^I_j$ where $j\in\{u,n\}$. To allow for the possibility of polarization in the bias of raters and notes, $f_u$ and $f_n$, we randomly assign each rater and note to one of two groups. The rater and note biases are then drawn independently from Gaussian distributions with parameters $\mu_{j}^F$ and $\sigma_j^F$ where $F \in \{+,-\}$ indicates the group. In particular, we set $\mu^+_j \geq 0$ and $\mu^-_j \leq 0$, such that members of the ``$+$'' group --- the \emph{positive bias group} --- tend to have bias $f>0$ and the ``$-$'' group --- the \emph{negative bias group} --- tend to have bias $f<0$. 

To establish the baseline efficacy of the Community Notes algorithm, we first tested its accuracy against populations of 2,000 notes and 2,000 raters, in which every rater gives a rating to every note. Note and rater parameters were drawn from unpolarized distributions, with $\mu_j^+ = \mu_j^{-} = 0$, $\mu^I_j=0.25$, and $\sigma^+_j = \sigma^-_j = \sigma^I_j = 0.25$ for $j \in \{u,n\}$ (see \SI Section~\ref{sec:si-distr-params}). We find that the Community Notes algorithm is highly accurate at classifying notes for this dataset, with suppression rate 0.013, pollution rate 0.070, waste rate 0.005, and infiltration rate 0.027.

Next we explored the accuracy of the Community Notes algorithm for simulated datasets of 20,000 notes and 10,750 raters, as described in \nameref{sec:methods}. We set $\mu^I_n = \mu^I_u = 0.25$ and $\sigma^I_n = 0.5$, which reproduces the empirical distribution of note ratings (see \nameref{sec:methods}). We then characterized the accuracy of the Community Notes algorithm as a function of three key features of the underlying rater and note parameter distributions: variability in rater friendliness, polarization of notes and raters, and degree of preference of raters for rating notes that share their bias (in-group bias). \emph{Variability in rater friendliness} is measured by the standard deviation parameter, $\sigma^I_u$. We define the \emph{note polarization} as the distance between the average bias of the two groups, i.e., $\rho_n = \frac{1}{2} \mu^+_n - \frac{1}{2} \mu^-_n \in [0,1]$. Similarly we define the \emph{rater polarization} as $\rho_u = \frac{1}{2} \mu^+_u - \frac{1}{2} \mu^-_u \in [0,1]$. We assume throughout that the groups are of equal size. Finally, we define the \emph{in-group bias} $E_h = \frac{2e_h}{E} - 1 \in [-1,1]$, where $E$ is the total number of ratings and $e_h$ is the number of ratings such that rater and note have the same bias, i.e., raters with $f_u>0$ rating notes with $f_n>0$ and vice versa (see \nameref{sec:methods}). 

\begin{figure*}[ht]
\centering
\includegraphics[width=0.6\textwidth]{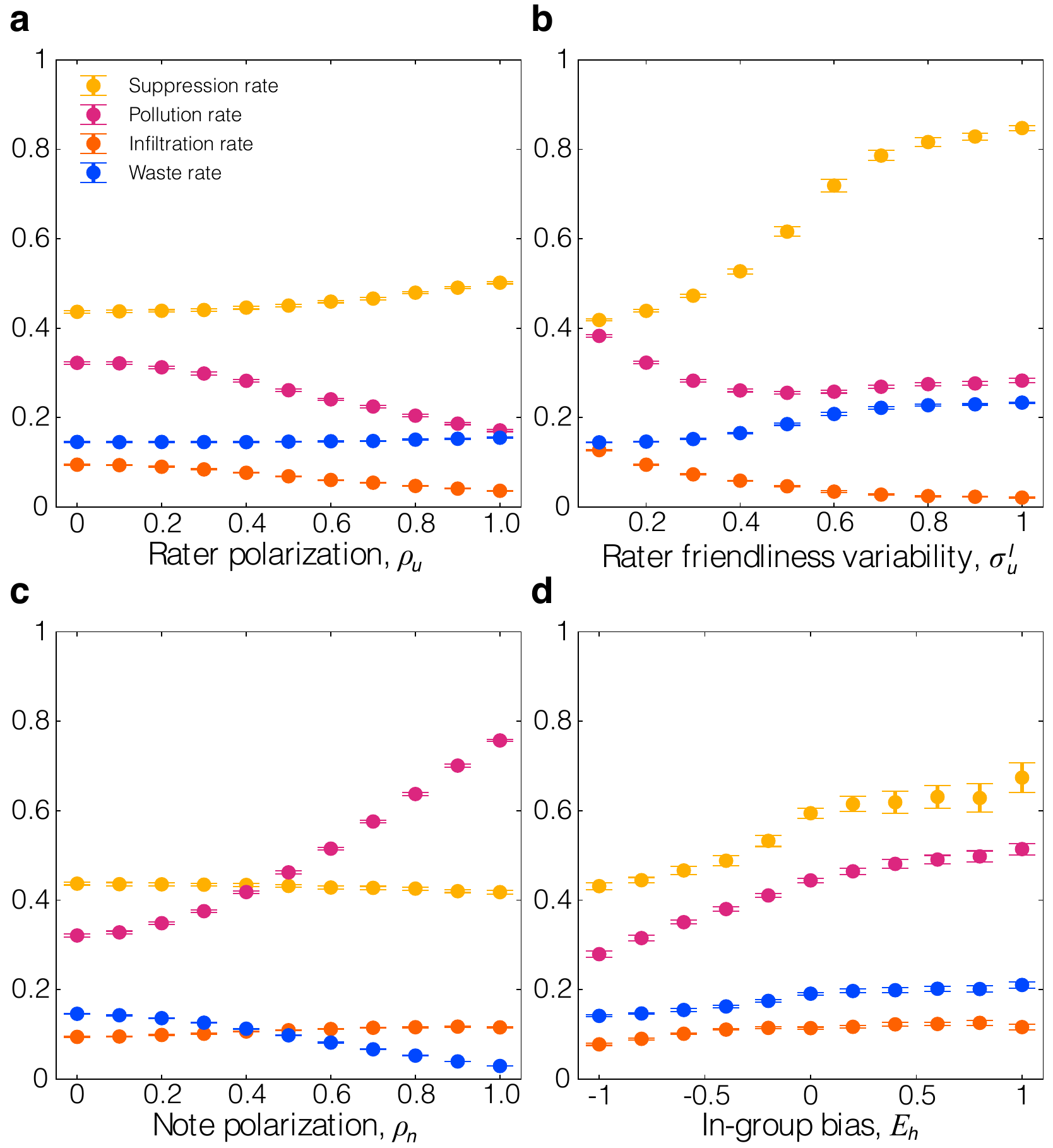}
\caption{\textbf{Varying the distributions of raters and notes.} We systematically varied the underlying distributions of note parameters, $i_n$ and $f_n$, and rater parameters, $i_u$ and $f_u$, in populations where all raters are honest.
(a)~Effects of rater polarization $\rho_u$ with unpolarized notes. (b)~Effects of variability of rater friendliness, as measured by standard deviation $\sigma^I_u$, in unpolarized populations. (c)~Effects of note polarization $\rho_n$ with unpolarized raters. (d)~Effects of in-group bias $E_h$ when choosing which notes to rate in unpolarized populations. Plots show the mean and standard errors across 50 replicate datasets of 20,000 notes and 10,750 raters, with the distribution of ratings chosen to replicate the real-world distribution (see \nameref{sec:methods}). In all cases we set $\mu^I_n = \mu^I_u = 0.25$ and $\sigma^I_n = 0.5$, which reproduces the empirical frequency of HELPFUL ratings.
}
\label{fig:variability}
\end{figure*}

We find that
the infiltration rate and waste rate remain relatively low across all conditions, rarely reaching values above 20\%. 
In contrast, the Community Notes algorithm tends to have an elevated suppression rate and pollution rate, with values of the suppression rate consistently above 40\% across all conditions (Figure~\ref{fig:variability}). 
Both sources of error are highly sensitive to the underlying rater and note distributions. The suppression rate reaches levels above 80\% (i.e., more than 80\% of helpful notes go unpublished) when raters have high variability in their tendency to rate notes as helpful (Figure~\ref{fig:variability}b). This means that a large number of misleading posts remain unlabeled, even though helpful notes are available. 
Similarly for the pollution rate, almost 80\% of published notes are in fact unhelpful when note polarization is high (Figure~\ref{fig:variability}c). Because social media notes tend to be polarized, this scenario is likely and could severely compromise the reliability of Community Notes. 

In-group bias and variability in rater friendliness drive higher suppression rates, while note polarization and in-group bias lead to high pollution rates. 
Perhaps more surprisingly, high levels of rater polarization or friendliness variability lead to lower pollution rates. Out-group bias, in which raters prefer to rate notes with the opposite bias ($E_h=-1$, i.e., raters with $f_u>0$ tend to rate notes with $f_n<0$ and vice versa), reduces both suppression and pollution rates (Figure~\ref{fig:variability}d).
This is consistent with the idea that Community Notes functions as a bridging algorithm \cite{ovadya2023bridging}, i.e., it works better when raters are drawn from across polarized divides. However, the suppression rate remains high even in the best-case scenario. 

\subsection*{Vulnerability to indiscriminate bad raters}

Next we assess the vulnerability of the Community Notes algorithm to ``bad raters.'' We conceive of bad raters as volunteers who seek to suppress the publication of helpful notes by down-rating them. This may take two qualitatively different forms: \textit{indiscriminate} bad raters who aim to disrupt the functioning of the Community Notes algorithm in general, by down-rating any notes that they perceive to be helpful; \textit{coordinated} bad raters who target and suppress the publication of notes with a particular bias, by down-rating only those they perceive to be helpful and to have either $f_n>0$ or $f_n<0$. 

Indiscriminate bad raters behave like normal, honest raters for any note that is not helpful enough to be published ($i_n<0.4$). If a note is helpful enough to be published ($i_n \geq 0.4$), bad raters treat it like an unhelpful note with a fixed probability, called the \textit{behavior rate} (see \nameref{sec:methods} for details).

To assess the impact of indiscriminate bad raters on the performance of the Community Notes algorithm, we systematically varied both the proportion of bad raters among all raters, as well as their bad behavior rate. As in the previous section we generated simulated datasets of 20,000 notes and 10,750 raters, where raters were randomly assigned to be bad raters (see \nameref{sec:methods}). For the underlying distribution of rater and note parameters, we set $\mu^I_n=\mu^I_u=0.25$, $\sigma^I_n=0.5$, and $\sigma^I_u=0.2$, which reproduces the empirical distribution of note ratings. For the rater and note bias, we set $\mu_j^+=\mu_j^{-}=0$ and $\sigma^+_j=\sigma^-_j=0.25$, so that there is no rater or note polarization ($\rho_u=\rho_n=0$). Finally we assume no in-group bias in the ratings ($E_h=0$). These parameter choices result in relatively low suppression and pollution rates when bad raters are absent, as shown in Figure~\ref{fig:variability}.

\begin{figure*}[ht]
\centering
\includegraphics[width=1.0\textwidth]{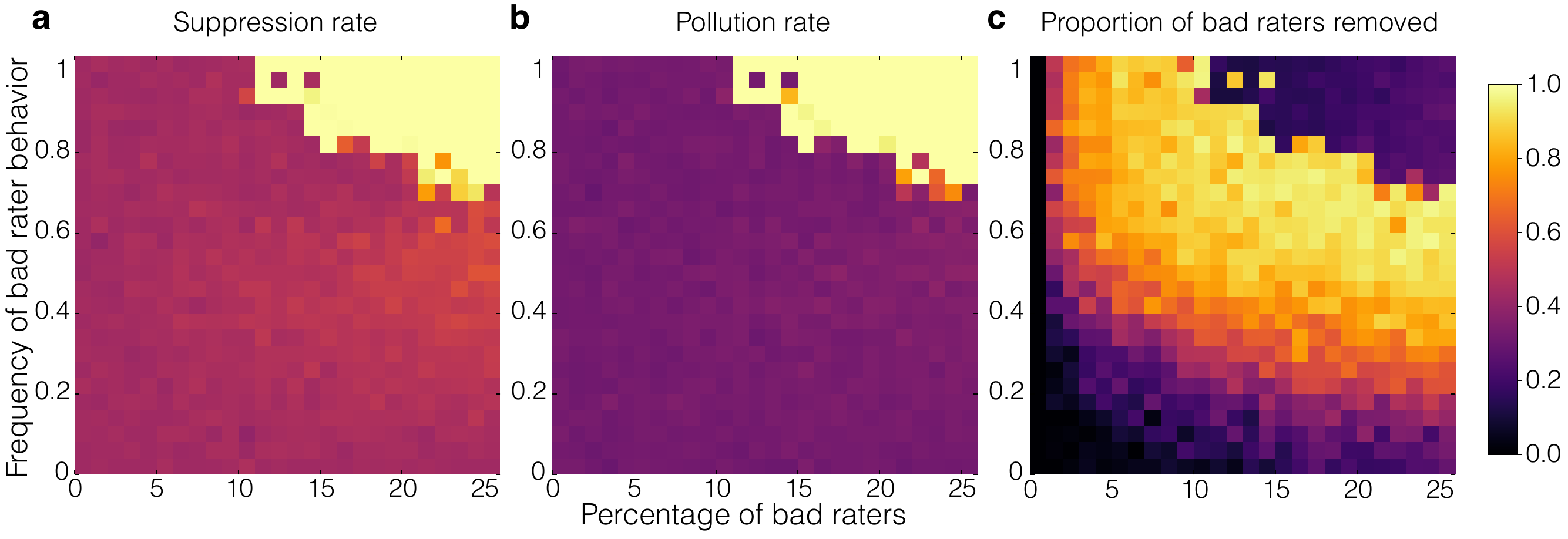}
\caption{\textbf{Indiscriminate bad raters}. We systematically varied the percentage of indiscriminate bad raters and the frequency with which they mis-rate notes they perceive as helpful. (a)~The suppression rate remains relatively low until the percentage of bad raters reaches 10--15\%, along with a frequency of bad rater behavior around 0.8, after which the suppression rate reaches 100\%, indicating that \emph{all} truly helpful notes go unpublished. (b)~A similar pattern occurs for the pollution rate, indicating that \emph{none} of the published notes are truly helpful. (c)~The helpfulness filter
tends to successfully remove bad raters when there are sufficiently few ($<10\%$) and/or the frequency of bad rater behavior is sufficiently low ($<0.8$). However the filter breaks down at the same time as the suppression and pollution rate approach 100\%. 
Results shown are for a single replicate with 20,000 notes and 10,750 raters, for each choice of bad rater percentage and frequency of bad rater frequency. Parameters are as described in Figure~\ref{fig:variability} and the main text, for unpolarized populations ($\rho_n=\rho_u=0$), and with no in-group bias ($E_h=0$). 
}
\label{fig:nonbiased-bad}
\end{figure*}

Figure~\ref{fig:nonbiased-bad} shows the impact of systematically varying the percentage of bad rater and frequency of bad behavior rate on the suppression and pollution rates of the Community Notes algorithm. We observe a critical transition whereby the Community Notes algorithm effectively breaks down, with both suppression rate and pollution rate reaching 100\% when approximately 12\% of raters are bad and consistently display bad behavior, or when more than 20\% of raters are bad but they only use bad behavior around 70\% of the time. This breakdown corresponds to a situation in which \emph{all} genuinely helpful notes go unpublished and \emph{all} published notes are in fact unhelpful. In other words, a small minority of bad raters, who do not coordinate in any way, but independently down-rate notes they think are likely to be published, can render the Community Notes algorithm totally non-functional. In contrast, the effect of bad raters on the infiltration and waste rate is minimal (Figure~\ref{fig:si-nonbiased-bad-errors}).

We can explain the breakdown of the Community Notes algorithm by looking at the efficacy of the helpfulness filter, which removes raters who consistently give inaccurate ratings compared to the consensus (see \nameref{sec:methods}). Figure~\ref{fig:nonbiased-bad}c shows  the proportion of bad raters removed by the Community Notes algorithm. 
When bad raters and/or bad behavior are rare, some are missed but this has little effect on error rates. 
As they become more common, the filter starts working well.  
However, the filter becomes ineffective at removing bad raters once they reach approximately 12\% of the rater population. 
Above that threshold, the bad raters are able to shift the consensus of the overall population of raters such that they are no longer labeled as unhelpful by the filter.

\subsection*{Vulnerability to coordinated bad raters}

While indiscriminate bad raters have a devastating impact on the Community Notes algorithm, they suppress notes regardless of bias. An even more important question is whether bad raters can effectively coordinate to suppress notes with a particular bias only. This would correspond to a situation in which, for example, a subset of raters with a particular political leaning tried to prevent helpful notes from being attached to posts from their own side.

\begin{figure*}[ht]
\centering
\includegraphics[width=1.0\textwidth]{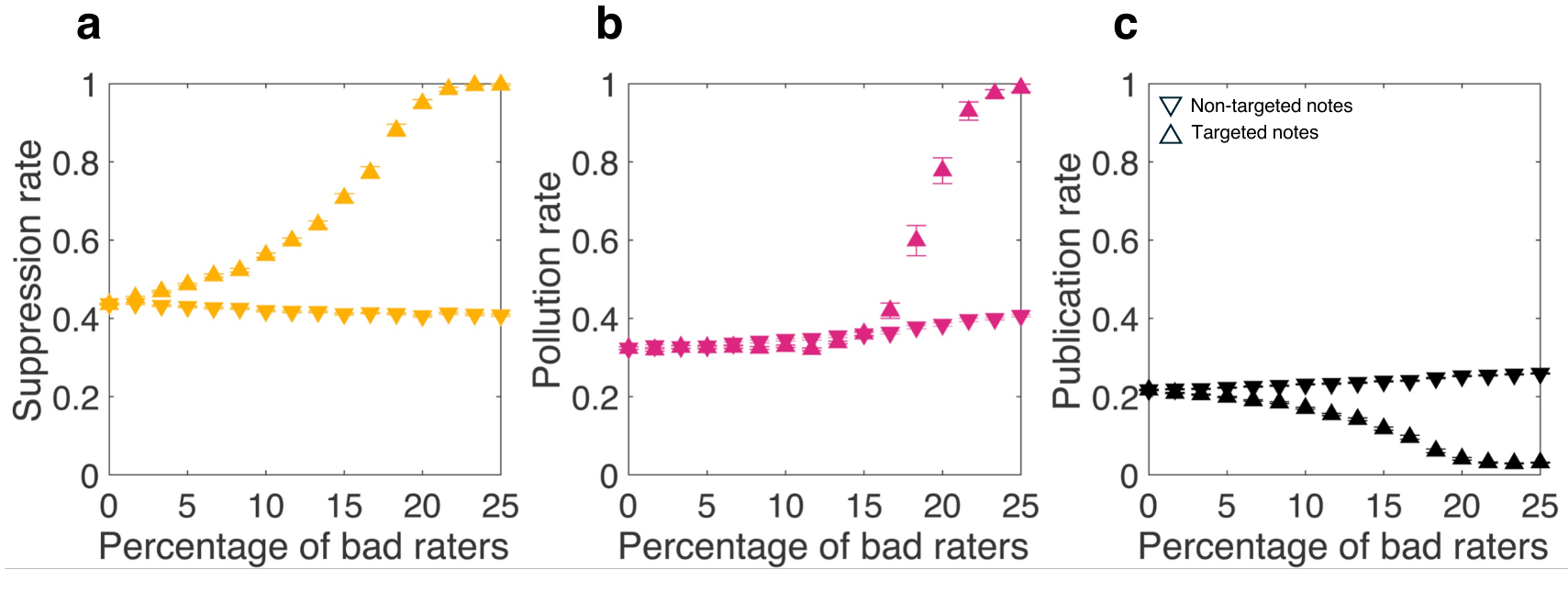}
\caption{\textbf{Coordinated bad raters}.  We systematically varied the percentage of coordinated bad raters, who mis-rate notes they perceive as helpful only if those notes belong to a targeted group. Here the frequency of bad rater behavior is set to one (we explore the more general case in \SI). We randomly set the targeted group to be either notes with $f_n<0$ or notes with $f_n\geq0$ for each replicate dataset. (a)~The suppression rate diverges for targeted and non-targeted notes once the percentage of bad raters reaches approximately 5\%, and the suppression rate for targeted notes reaches 100\% once the percentage of bad raters reaches approximately 20\%, while the suppression rate of non-targeted notes remains unchanged. (b)~The pollution rate for targeted and non-targeted notes diverges once the percentage of bad raters reaches approximately 15\% and reaches 100\% once the percentage of bad raters reaches approximately 25\%, while the pollution rate for non-targeted notes again remains largely unchanged. (c)~The publication rate of targeted notes declines from around 20\% when there are no bad raters to around 1\% when bad raters exceed 20\%, while the publication rate of non-targeted notes remains largely unchanged.
Results shown are the mean and standard error for 100 replicates with 20,000 notes and 10,750 raters, for each choice of bad rater percentage. Parameters are as described in Figure~\ref{fig:variability} and the main text, for unpolarized populations ($\rho_n=\rho_u=0$) and with no in-group bias ($E_h=0$).  
}
\label{fig:biased-bad}
\end{figure*}

To capture such a scenario, we implement bad raters in the same way as above, with the additional constraint that they only down-rate notes that are helpful enough to be published ($i_n \geq 0.4$) and belong to a target group, which we assume means having a specific bias (e.g. $f_n<0$; see \nameref{sec:methods}). Figure~\ref{fig:biased-bad} shows the impact of such coordinated bad raters on the functioning of the Community Notes algorithm. Under the same baseline scenario, the frequency of bad raters required to cause the Community Notes algorithm to break down is higher than that in the case of indiscriminate bad raters. Suppression and pollution rates reach 100\% only when the fraction of coordinated bad raters reaches approximately 20\%. However, the nature of the breakdown is completely asymmetric: \emph{all} targeted notes that should be published are suppressed. Yet notes that are not targeted are not impacted at all by the presence of bad raters. 
In other words, a coordinated group of bad raters can effectively manipulate the Community Notes algorithm to stop publishing helpful notes with a particular bias. 
This analysis assumes that bad raters consistently follow bad behavior; the effects of bad raters on infiltration and waste rates are minimal, as shown in \ref{fig:si-biasedbad-errors-x}. We also explore a range of bad rater proportions and different frequencies of bad behavior, presented in Figure~\ref{fig:si-biasedbad-filter-true}.

While the deleterious impact of indiscriminate bad raters may be hard to detect unless the quality of the published notes is independently verified \cite{wojcik2022birdwatch}, the impact of coordinated bad raters can be detected more easily, since they also suppress the overall number of published notes with the targeted bias.
Figure~\ref{fig:biased-bad}c shows that  
as the number of coordinated bad raters increases, the number of targeted notes published declines from around 20\% 
to around 1\%. This is in contrast to the publication rate for notes that are not targeted for suppression, which increases slightly. 

\begin{figure*}[ht]
\centerline{\includegraphics[width=0.6\textwidth]{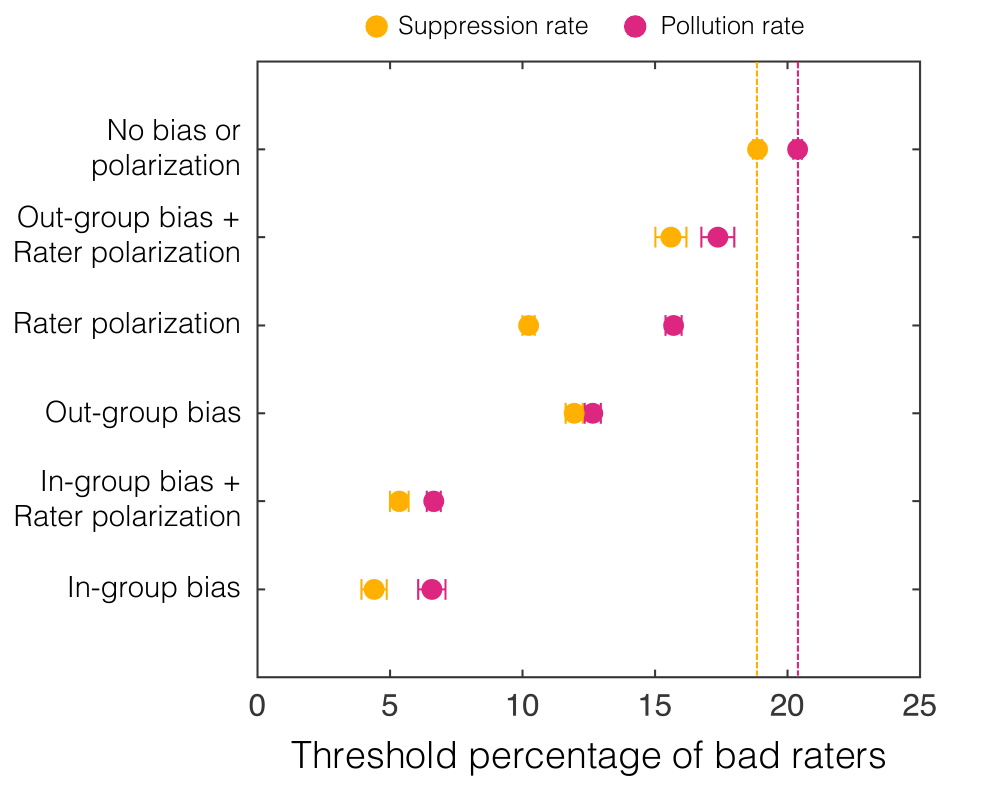}}
\caption{\textbf{Interaction between rater bias and critical percentage of bad raters}. We calculated the percentage of coordinated bad raters required to raise the suppression rate (yellow) and pollution rate (pink) to 90\% under different experimental conditions. The frequency of bad rater behavior is set to one. 
For all raters, we considered cases with out-group bias ($E_h=-1$), in-group bias ($E_h=1$), and neither ($E_h=0$). We combined these cases with the presence of rater polarization ($\rho_u=1$) or its absence ($\rho_u=0$). Results shown are the means and standard errors across 100 replicates with 20,000 notes and 10,750 raters. 
Parameters are as described in Figure~\ref{fig:variability} and the main text, for populations without note polarization ($\rho_n=0$).}
\label{fig:homophily-x-badpct} 
\end{figure*}

\subsection*{Interaction between bias and manipulation}

We have shown that different types of bias by honest raters, such as polarization and in-group bias, can suppress the publication of a majority of helpful notes (Figure~\ref{fig:variability}). Bad raters who actively seek to down-rate helpful notes can have even worse effects (Figure~\ref{fig:nonbiased-bad}-\ref{fig:biased-bad}). In reality, both problems can exist side-by-side. In particular, the presence of polarization and selection bias among honest raters may make it easier for bad raters to degrade the output of the Community Notes algorithm. 
To test this, we calculate the threshold percentage of coordinated bad raters required to generate either a suppression rate or a pollution rate of 90\%. 
Figure~\ref{fig:homophily-x-badpct} shows that, depending on the interaction between polarization and in/out-group bias by all raters, the critical percentage of bad raters for suppression varies from a minimum of 5\% to a maximum of 19\%, with a similar pattern for pollution (7--20\%).

Most strikingly, both in-group and out-group bias among the whole population of raters tends to make it easier for bad raters to disrupt Community Notes. 
Rater polarization also makes the algorithm more vulnerable to manipulation by bad raters. Thus the Community Notes algorithm is most robust when presented with un-polarized raters who tend to rate notes without regard for their bias. But even in this ideal case, 20\% of bad raters can still break down the system.

\subsection*{Impact of errors on the helpfulness of notes}

While we have focused on error rates, we have not accounted for the severity of the errors. Even if the true helpfulness of published notes falls short of the threshold for publication set by the algorithm, generally helpful notes may still be published.
To determine whether this is the case, we calculated the excess (true) helpfulness of notes that are published compared to those that should be published, and that of notes that are not published to those that should not be published (see \nameref{sec:methods}). 
We also make similar calculations for the excess absolute bias (see \SI Section~\ref{sec:si-excess-helpfulness-bias} and Figures~\ref{fig:si-biasedbad-bias-avg}-\ref{fig:si-biasedbad-scatter}).

Table~\ref{table:excess} shows that, in the baseline condition of no bad raters, there is only a small effect of errors on the helpfulness of published notes. 
In a similar condition but with 25\% coordinated bad raters, the helpfulness of targeted notes that are published is reduced by more than half, while the helpfulness of published non-targeted notes is reduced by only a few percentage points. In contrast, unpublished targeted notes are almost 235\% more helpful than expected, while unpublished non-targeted notes are about 30\% more helpful. 
Thus bad raters not only cause the Community Notes algorithm to make errors; they cause it to make errors that materially reduce the helpfulness of published notes. A full breakdown of the excess bias across experimental conditions can be found in \SI Section~\ref{sec:si-excess-helpfulness-bias}.
We also calculated the Pearson correlation between the true and inferred helpfulness of notes, $i_n$ and $\hat{i}_n$, and between the true and inferred note bias, $f_n$ and $\hat{f}_{n}$. The correlations reiterate our findings above, based on error rates, that the algorithm becomes less accurate in the presence of bad raters (Figure~\ref{fig:si-biasedbad-corr}).

\begin{table}
\caption{\textbf{Excess helpfulness for published and unpublished notes.}
We compare the excess helpfulness of targeted and non-targeted notes when coordinated bad raters make up 25\% of the population, to the excess helpfulness of the same groups of notes when bad raters are absent. We assume no in-group or out-group bias and no polarization in both cases. There is no significant difference between the excess helpfulness of targeted notes and that of non-targeted notes, for both published and unpublished notes when bad raters are absent. Coordinated bad raters reduce the helpfulness of targeted notes that are published and increase the helpfulness of targeted notes that are unpublished, compared to non-targeted notes.
}
\begin{tabular}{lcccc}
\hline
 & \multicolumn{2}{c}{0\% Bad Raters} & \multicolumn{2}{c}{25\% Bad Raters} \\
\cline{2-5}
 & Published (\%) & Unpublished (\%) & Published (\%) & Unpublished (\%) \\
\hline
Targeted notes & $-1.22 \pm 0.29$ & $53.17 \pm 1.73$ & $-54.90 \pm 1.07$ & $234.99 \pm 3.65$ \\
Non-targeted notes & $-1.28 \pm 0.27$ & $53.78 \pm 1.88$ & $-7.76 \pm 0.38$ & $30.60 \pm 2.00$ \\
\hline
\end{tabular}
\label{table:excess}
\end{table}


\section*{Discussion}

Social media platforms seek to make users aware of potential misinformation using moderation approaches that work at scale and that avoid the appearance or reality of censorship. 
Crowdsourced accuracy ratings provide an alluring potential solution to this problem, but are ineffective when raters are not independent or are influenced by hubs in a social network \cite{Centola2017Crowds}. 
Community moderation is thus vulnerable to high levels of rater bias, such as polarized social influence or active manipulation of the process. 
Unfortunately, assessing the extent of these vulnerabilities is difficult because it requires knowledge about the intentions and biases of volunteer raters.

Our approach gets around this problem by simulating data that recapitulates the number and distribution of ratings in the empirical data. These realistic simulations provide full knowledge of rater bias and decision-making processes, allowing us to systematically evaluate the performance of the Community Notes algorithm under different scenarios.  
We exposed the algorithm to over fifteen thousand simulated datasets, varying the underlying distributions of both rater and note biases, and introducing bad raters who actively seek to suppress helpful notes. 
We found that the algorithm generally fails to accurately infer the bias and helpfulness of notes and raters, resulting in high error rates when deciding which notes to publish and which to leave unpublished. 

The proportions of genuinely unhelpful notes that are published, as well as helpful notes that remain unpublished tend to be below 20\% across most conditions. 
Viewed in isolation, these quantities would give the impression that the algorithm is accurate and robust. 
However, over 40\% of genuinely helpful notes go unpublished, and the proportion of published notes that are in fact unhelpful can also be very high. 
In other words, under reasonable assumptions about rater and note biases, the algorithm fails to surface many helpful notes, and the overall quality of published notes can be low. 

Both indiscriminate bad raters, who seek to suppress helpful notes regardless of note bias, and coordinated bad raters, who target helpful notes with a particular bias, can push the suppression and pollution rates up to values close to 100\%. In these scenarios, the algorithm publishes almost no truly helpful notes, and instead surfaces mostly unhelpful notes. 
Most strikingly, coordinated bad raters can completely suppress targeted helpful notes while leaving non-targeted notes largely unaffected. 
As little as 5--20\% of bad raters can completely degrade the Community Notes algorithm, depending on the degree of polarization and in-group bias among the general population of raters. 
The helpfulness filter, designed to remove low-quality raters, is unable to identify these raters once they reach such critical proportions. 

Successful suppression of targeted notes (90\% or greater) coincides with both low levels of helpfulness among published notes (pollution rate also at or above 90\%) and a decline in the overall number of targeted notes being published. High error rates also result in the Community Notes algorithm becoming less accurate at estimating note bias (see \SI Section~\ref{sec:si-excess-helpfulness-bias}). 
The suppression and pollution rates in our baseline condition --- no bad raters, polarization, or in-group bias --- are lower (30--40\%) and correspond to the publication of notes whose helpfulness is high (Table~\ref{table:excess} and \SI Section~\ref{sec:si-excess-helpfulness-bias}).

Our approach is necessarily limited in its ability to capture the decision-making process of real raters, especially their underlying biases. Our goal was to recapitulate as closely as possible the assumptions of the Community Notes algorithm. While a single number may be sufficient to capture a country's main political dimension \cite{bouchaud2025algorithmic}, bias in general has many other potential dimensions, for example those corresponding to race, gender, age, and socioeconomic status \cite{doi:10.1073/pnas.2102140118}. However, we made the same assumption as the Community Notes algorithm that rater and note bias are each captured by a single number, $f_u$ and $f_n$. 
Our simulated data therefore provides the most basic test possible that we would expect the Community Notes algorithm to pass. The fact that it often fails to do so suggests that its ability to deal with factors such as multi-dimensional identity and bias is severely limited.   

While previous studies have suggested that currently published Community Notes are generally accurate \cite{allen2021scaling, BARBERA2024103792} and effective \cite{drolsbach_diffusion_2023, kankham_community_2024, slaughter2025community}, our analysis raises two concerns regarding these observations. First, the high suppression rates in our experiments suggest that many genuinely helpful notes go unpublished, in agreement with empirical studies \cite{ccdh, mohammadi2025birdwatch}. 
This not only means that misinformation remains unlabeled, but also that the authors of these notes receive no recognition for their work, which could gradually erode the volunteer base over time \cite{Alexios25}. Second, the algorithm is highly vulnerable to coordinated bad raters, especially if the volunteer base becomes increasingly polarized or focused on in-group ratings. Recent moves to integrate LLMs into Community Notes \cite{Li_CN+LLM_2025} to help writers be more efficient might also increase the vulnerability of the algorithm to this kind of coordination. Furthermore, LLMs are themselves vulnerable to bias \cite{Yang2023LLM-rate-news} and manipulation \cite{souly2025poisoningattacksllmsrequire}.

The key question in response to our findings is what can be done to mitigate the problems we have identified. While we do not have a silver bullet, we can make a few promising observations. 
We find that in-group and out-group bias can both increase the vulnerability of the algorithm to bad raters, suggesting that norms encouraging raters to engage with all types of note, regardless of bias, can improve the resilience of Community Notes to coordinated attacks. This is consistent with the view that Community Notes works best as a bridging algorithm \cite{ovadya2023bridging}. 
We also find that the number of published notes declines in the group targeted by bad actors. A large change in publication rate among a targeted group could potentially serve as a signal of coordinated suppression. However, identifying such a signal requires notes to be correctly assigned to the targeted group. If targeting is based on note bias (targeting of notes with $f_n<0$, for example), categorization of notes becomes difficult, because the accuracy of the inferences made by the Community notes algorithm, $\hat{f}_n$, can be low (Figure~\ref{fig:si-biasedbad-bias-avg}). This highlights the need for some external validation mechanism for determining note bias in order to correctly label notes and thus defend against coordinated attacks. 
Finally, the algorithm is vulnerable to bad raters due to the breakdown of the helpfulness filter. This suggests that alternative approaches to filtering, such as focusing on raters who under-value the helpfulness of published notes, may be productive. 
Our approach using simulated data makes it possible to explore different types of filtering and different types of optimization, and thus to systematically improve the accuracy and resilience of crowdsourcing algorithms. 






\section*{Methods}
\label{sec:methods}

\subsection*{Community Notes code and data}

We used the open Community Notes code and data published on June 27, 2025 as the basis for our simulated data. The precise version we used can be found in our Github repository (see below).
This dataset contains 132,958,960 ratings on 1,869,267 notes rated by 1,153,498 raters. 

\subsection*{Generating simulated ratings}

According to the Community Notes algorithm \cite{wojcik2022birdwatch, cn-guide}, 
a note $n$ receives a rating $r_{un}$ from a rater $u$. Ratings take the value of either 1 (``HELPFUL''), 0.5 (``SOMEWHAT HELPFUL''), or 0 (``NOT HELPFUL''). The value is ``NULL'' if the rater has not rated the note. 
The algorithm then \emph{models} the interaction between raters and notes as
\(
\hat{r}_{un}=\hat{\mu}+\hat{i}_u+\hat{i}_n+\hat{f}_u\hat{f}_n
\)
and 
estimates the values of $\hat{i}_u$, $\hat{i}_n$, $\hat{f}_u$, $\hat{f}_n$ and $\hat{\mu}$ 
by minimizing via gradient descent the least-square error loss function
$
\sum_{r_{un}}(r_{un}-\hat{r}_{un})^2+\lambda_i(\hat{i}_u^2+\hat{i}_n^2+\hat{\mu}^2)+\lambda_f(\hat{f}_u^2+\hat{f}_n^2)
$
where $\lambda_i$ and $\lambda_f$ are regularization factors (set to $\lambda_i=0.15$ and $\lambda_f=0.03$ \cite{wojcik2022birdwatch}). 

The algorithm removes some ratings based on a ``helpfulness filter'', which excludes raters whose ratings consistently mis-align with the status decision of the Community Notes algorithm. In particular, at least two thirds of the ratings need to match the modeled HELPFUL or UNHELPFUL status. If a rater fails to meet this criterion, their ratings will be removed from the modeling processing. This filter aims to remove low quality raters \cite{wojcik2022birdwatch}.


We assume raters employ a softmax activation function to determine how to rate a note given the parameters $i_u$, $i_n$, $f_u$, $f_n$, and $\mu$.  
Note $n$ is rated HELPFUL by rater $u$ with probability
\begin{equation}
p_{H}=\frac{\exp[-\gamma(\frac{1}{2}-(\mu+i_u+i_n+f_uf_n))]}{1+\exp[-\gamma(\frac{1}{2}-(\mu+i_u+i_n+f_uf_n))]+\exp[\gamma(\frac{1}{2}-(\mu+i_u+i_n+f_uf_n))]}
\label{eq:1}
\end{equation}
and UNHELPFUL with probability
\begin{equation}
p_{\bar{H}}=\frac{\exp[\gamma(\frac{1}{2}-(\mu+i_u+i_n+f_uf_n))]}{1+\exp[-\gamma(\frac{1}{2}-(\mu+i_u+i_n+f_uf_n))]+\exp[\gamma(\frac{1}{2}-(\mu+i_u+i_n+f_uf_n))]}
\label{eq:2}
\end{equation}
while the probability of rating a note as SOMEWHAT HELPFUL is $p_{SH}=1-p_{H}-p_{\bar{H}}$. 
The global parameter $\gamma$ controls the amount of noise in the rating decision process. 

We calibrated the model parameters to reproduce the empirical proportions of ratings of each type. 
The Community Notes dataset made available by X has 59.6\% HELPFUL ratings, 37.4\% NOT HELPFUL ratings, and 3.0\% SOMEWHAT HELPFUL ratings.
We systematically varied the noise parameter $\gamma$ and the mean and variance of the distributions of note and rater bias and helpfulness in order to match 
the percentage of HELPFUL and NOT HELPFUL ratings to
the empirical distribution. We found that the distribution of note ratings is sensitive only to mean note helpfulness $\mu^I_n$, mean rater helpfulness $\mu^I_u$, and the noise parameter $\gamma$ (see \SI Section~\ref{sec:si-distr-params}), resulting in our default parameter choices $\mu^I_n=\mu^I_u=0.25$ and $\gamma=30$. There is some degree of flexibility in this choice, since more than one parameter combination reproduces the empirical distribution (see \SI Section~\ref{sec:si-distr-params}). Finally we set the global parameter $\mu=0.17$ to match that inferred by the Community Notes algorithm when run on the full empirical dataset.

Our simulated datasets comprise 20,000 notes and 10,750 raters, with 1,839,726 ratings distributed to match the empirical degree distribution of the rater-note network. To generate this, we constructed the empirical rater-note network, comprised of 1,869,267 notes and 1,153,498 raters, in which an edge connects a given note and rater if the rater provided a rating for that note. 

We then randomly sampled the degree of 20,000 notes from the empirical network, with the constraint that only notes with at least five ratings could be accepted. We then ``filled in'' the edges of the sampled notes by sequentially sampling the degree of single raters, with the constraint that only raters who have rated at least 10 notes could be accepted, and connecting each of the sampled rater edges at random to available notes, until all of the notes had the required number of ratings. This process resulted in 10,750 raters in total. The sampled network has a degree distribution that closely matches that of the empirical network (see Figure~\ref{fig:si-degree-distr}). 

When producing a simulated dataset, we randomly rewired $10^6$ pairs of edges of the sampled network to remove any potential bias due to the structure of a given rater-note network.


\subsection*{In- and out-group bias}

To generate realistic datasets, with features that match those observed in patterns of real rating behavior, we allow for the possibility of polarization in the distributions of note and rater bias, as described in the main text, as well as in-group or out-group bias~\cite{allen2022birds}, in which raters may tend to rate notes with the same bias as theirs (in-group bias), or with the opposite bias (out-group bias).

To generate datasets with in- or out-group bias, we first assign each rater and note to a group $+$, which tends to have $f_n>0$, or a group $-$, which tends to have $f_n<0$. We then rewire the edges of the rater-note network using the following rule: with probability $p$, a given rater connects to a note that belongs to their in-group. With probability $1-p$, they connect to a note that belongs to their out-group. A very large number of such rewiring events results in a network in which a given edge connects a note and rater of the same group with probability $p$. Thus if $p=1$, notes are only rated by raters of the same group (in-group bias), and if $p=0$, notes are rated by raters of the other group (out-group bias).

The expected number of edges connecting a note and rater of the same group (i.e., the same bias) is $\left<e\right>=pE$. Thus our bias measure $E_h=2e_h/E-1$ has expected value $\left<E_h\right>=2p-1$ for a given network generated this way. 

\subsection*{Bad raters}

In a given simulated dataset, each rater is randomly assigned to be bad with a fixed probability. 
A bad rater seeks to suppress each helpful note with probability determined by the bad behavior rate parameter. Otherwise, they behave like a normal (honest) rater. 

Indiscriminate bad behavior uses the following rule:
\begin{itemize}
    \item If $i_n<0.4$,  rate notes as HELPFUL with probability   
    $p_{H}$ as given in Eq.~\ref{eq:1},
    and UNHELPFUL with probability
    $p_{\bar{H}}$ as given in Eq.~\ref{eq:2}. 
    \item If $i_n\geq0.4$, rate notes as HELPFUL with probability $p^*_{H} = p_{\bar{H}}$ as given in Eq.~\ref{eq:2}
    and UNHELPFUL with probability $p^*_{\bar{H}}= p_{H}$ as given in Eq.~\ref{eq:1}. 
\end{itemize}
In other words, rate a likely helpful note according to a rule that swaps HELPFUL and UNHELPFUL labels.

Similarly for coordinated bad raters, we assume that  notes with either $f_n>0$ or $f_n<0$ are targeted. We randomize the bias of the notes targeted by the bad raters in any given simulated dataset. Let the bias of bad raters be $\phi=\{-1,1\}$. Coordinated bad behavior then uses the following rule:
\begin{itemize}
    \item If $i_n<0.4$ or $\phi f_n>0$,  rate notes as HELPFUL with probability   
    $p_{H}$ as given in Eq.~\ref{eq:1},
    and UNHELPFUL with probability
    $p_{\bar{H}}$ as given in Eq.~\ref{eq:2}. 
    \item If $i_n\geq0.4$ and $\phi f_n<0$, rate notes as HELPFUL with probability  
    $p^*_{H} =p_{\bar{H}}$ as given in Eq.~\ref{eq:2},
    and UNHELPFUL with probability
    $p^*_{\bar{H}}=p_{H}$ as given in Eq.~\ref{eq:1}. 
\end{itemize}
In other words, rate a likely helpful note with bias $\phi f_n<0$ according to a rule that swaps  HELPFUL and UNHELPFUL labels.


\subsection*{Assessing the accuracy of the Community Notes algorithm} 


In measuring the accuracy of the algorithm's inferences of $\hat{f}_n$ and $\hat{i}_n$, we are interested in four types of potential errors: Type I or II (false positives/negatives), depending on whether helpful/unhelpful notes are considered the positive/negative class.

For a given simulated dataset we calculate the number of notes that are both published by the Community Notes algorithm and \emph{in fact} helpful, $N_{ph}$. This corresponds to notes for which $i_n>0.4$, $|f_n|<0.5$, $\hat{i}_n>0.4$, and $|\hat{f}_n|<0.5$. Similarly the number of notes that are not published by the Community Notes algorithm, ($\hat{i}_n<0.4$ or $|\hat{f}_n|>0.5$), but are in fact helpful, ($i_n>0.4$ and $|f_n|<0.5$), is $N_{\bar{p}h}$. The number of notes that are published ($\hat{i}_n>0.4$ and $|\hat{f}_n|<0.5$) but are in fact unhelpful ($i_n<0.4$ or $|f_n|>0.5$) is $N_{p\bar{h}}$. The number of notes that are not published ($\hat{i}_n<0.4$ or $|\hat{f}_n|>0.5$) and are in fact unhelpful ($i_n<0.4$ or $|f_n|>0.5$) is $N_{\bar{p}\bar{h}}$.

To avoid confusion, we adopt descriptive names for the four types of error rates relevant to the context of publishing Community Notes: 

\begin{itemize}
\item \emph{Pollution rate} $\frac{N_{p\bar{h}}}{N_{p\bar{h}}+N_{ph}}$: the probability that a published note is in fact unhelpful. 
\item \emph{Suppression rate} $\frac{N_{\bar{p}h}}{N_{\bar{p}h}+N_{ph}}$: the probability that a genuinely helpful note is not published. 
\item \emph{Infiltration rate} $\frac{N_{p\bar{h}}}{N_{\bar{p}\bar{h}}+N_{p\bar{h}}}$: the probability that a genuinely unhelpful note is published. 
\item \emph{Waste rate} $\frac{N_{\bar{p} h}}{N_{\bar{p}h}+N_{\bar{p}\bar{h}}}$: the probability that an unpublished note is in fact helpful. 
\end{itemize}

To characterize the effect of errors on the quality of published notes, we calculated the true helpfulness $h_p$ and $h_{\bar{p}}$ of \emph{published} notes, i.e., the notes that meet the threshold for publication according to the algorithm, and of \emph{unpublished} notes, respectively (see \SI Section~\ref{sec:si-excess-helpfulness-bias}).
We compared these to the true helpfulness $H_p$ and $H_{\bar{p}}$ of \emph{publishable} notes, i.e., the notes whose true parameters meet the threshold for publication, and \emph{unpublishable} notes, respectively. We then calculated the \textit{excess helpfulness} of published notes compared to publishable notes, $\frac{h_p}{H_p}-1$, and of unpublished notes compared to unpublishable notes, $\frac{h_{\bar{p}}}{H_{\bar{p}}}-1$. Finally we define the \emph{publication rate}, $\frac{N_{p\bar{h}} + N_{ph}}{N_{p\bar{h}} + N_{ph}+N_{\bar{p}\bar{h}} + N_{\bar{p}h}}$. 

\section*{Data Availability Statement} 

Code and data to reproduce the results are available at \url{https://github.com/osome-iu/communitynotes-manipulation}. 

\section*{Acknowledgments}

This work was supported in part by the Swiss National Science Foundation (Sinergia grant CRSII5\_209250), NSF (grant BCS-2444659), and the Knight Foundation.

\clearpage

\setcounter{figure}{0} 
\renewcommand{\thefigure}{\arabic{figure}} 

\def\thefootnote{\arabic{footnote}}

\setcounter{page}{1}
\renewcommand\thepage{S\arabic{page}}

\begin{centering}
{\huge
\textbf{Supplementary Materials}
}
\end{centering}

\tableofcontents
\newpage

\clearpage

\appendix


\counterwithin{figure}{section}
\counterwithin{table}{section}

\counterwithout{figure}{section}
\counterwithout{table}{section}

\renewcommand{\thefigure}{S\arabic{figure}}
\renewcommand{\thetable}{S\arabic{table}}

\setcounter{figure}{0}
\setcounter{table}{0}


\section{Supplementary Methods}

In this section we provide additional information on our simulation methods.

\subsection{Generating the rater-note network}
\label{sec:si-network}

\begin{figure*}[bh]
\centering
\includegraphics[width=0.8\textwidth]{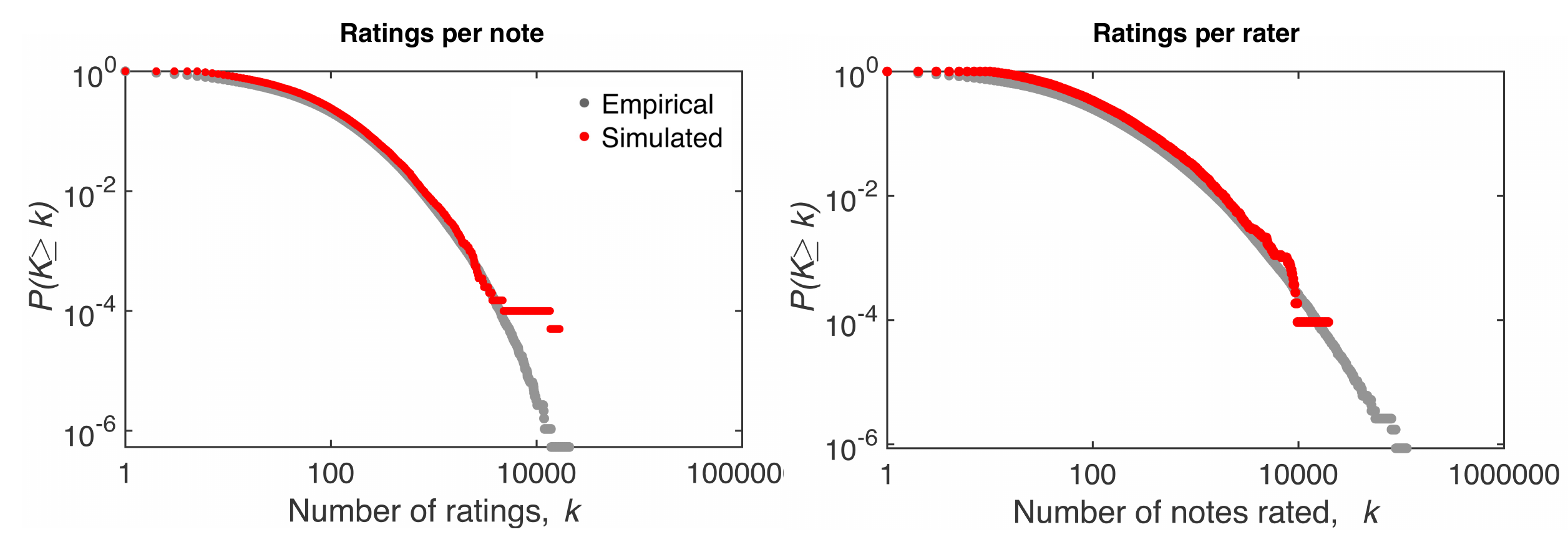}
\caption{\textbf{Complementary cumulative distributions of degree in empirical and simulated networks.} 
The Complementary Cumulative Distributions Function (CCDF) plots the probability that a randomly selected node has degree $k$ or higher.
(Left) CCDF for the empirical distribution of ratings per note for 1,869,267 notes (gray) and for a sampled  seed distribution of 20,000 notes (red). 
(Right) CCDF for the empirical distribution of ratings per rater for 1,153,498 notes (gray) and for a sampled  seed distribution of 10,750 raters (red).}
\label{fig:si-degree-distr}
\end{figure*}

We generate rater-note networks as described in \nameref{sec:methods}, by sampling from the empirical network downloaded from Community Notes. This provides us with a ``seed network'' of the required size (20,000 notes and 10,750 users unless noted otherwise) that forms the basis for each simulated dataset. When generating a given dataset, we rewire the entire network, corresponding to $10^6$ pairs of individual rewiring events. To do this we select pairs of edges and swapping their target notes, with the constraint that each rater can rate each note only once. In-group bias is implemented by ensuring that, when rewiring an edge, the rater and note groups are the same with probability $p$. As a result of this process, all of our networks have the same degree distribution, which closely matches that of the empirical distribution, as shown in Figure~\ref{fig:si-degree-distr}.

\subsection{Distributions of rater and note parameters}
\label{sec:si-distr-params}

We drew true rater and note parameters, $i_u$, $f_u$, $i_n$ and $f_n$ independently from gaussian distributions as described in the main text and results. We set the global parameter $\mu=0.17$ to match that inferred by the Community Notes algorithm run on the full empirical dataset. Since our model assumes that rating decisions, $r_{un}$, are the output of a softmax function (Eqs.~1--2), we also have a noise parameter $\gamma$. 

This leaves us with the following free parameters that must be chosen to generate a given dataset: $\mu^I_u$, $\mu^I_n$, $\sigma^I_u$, $\sigma^I_n$, $\mu^+_u$, $\mu^+_n$, $\mu^-_u$, $\mu^-_n$, $\sigma^+_u$, $\sigma^+_n$, $\sigma^-_u$, $\sigma^-_n$, and $\gamma$ (i.e. the mean and standard deviations of six gaussian distributions and noise). We simplified matters by setting $\mu^I_u=\mu^I_n$, $\mu^+_u=-\mu^-_u$, $\mu^+_n=-\mu^-_n$, $\sigma^+_u=\sigma^-_u$, and $\sigma^+_n=\sigma^-_n$. This reflects assumptions that the helpfulness of users who write and rate notes is the same, and that the distribution of bias is the same for each group. Eight parameters that must therefore be chosen. Our goal is to preserve the empirical proportions of note ratings --- 59.6\% HELPFUL, 37.4\% NOT HELPFUL, and 3.0\% SOMEWHAT HELPFUL --- while allowing us to model polarization, in-group bias, and rater friendliness variability.

\begin{figure*}
\centering
\includegraphics[width=0.9\textwidth]{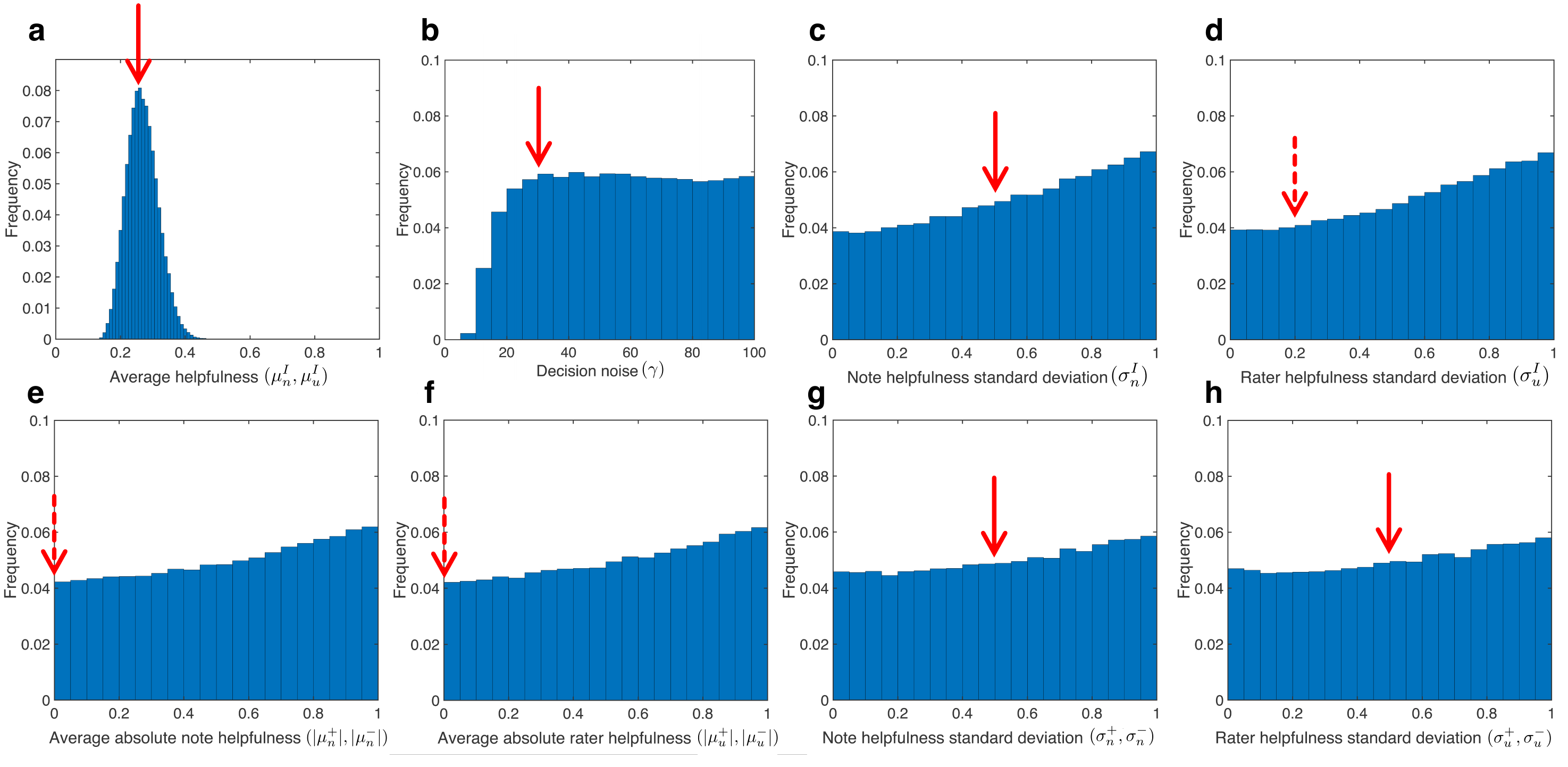}
\caption{\textbf{Distribution parameters for notes and raters.} We drew $10^7$ parameter combinations (see text) and retained those that produced rating proportions within 2\% of the empirical distribution. The resulting distributions of parameters are shown. (a)~The mean rater friendliness and note helpfulness, $\mu^I_u=\mu^I_r$ is constrained to values 0.2--0.4. (b)~The noise parameter $\gamma$ is constrained to values $>10$. (c--h)~All other parameter are largely unconstrained. Arrows indicate the default parameters, used in our baseline condition. Solid arrows indicate parameters that remain fixed in the main text, dashed arrows indicate values for the baseline condition (no polarization, low rater friendliness variability), which are varied in Figure~\ref{fig:variability} in main text.}
\label{fig:si-distributional-params}
\end{figure*}

We drew $10^7$ parameter combinations with means constrained to $\mu\in[-1,1]$, standard deviations constrained to $\sigma\in[0,1]$, and the noise parameter constrained to $\gamma\in[0,100]$. We then retained any parameter combination that satisfied the following condition: $(\left<p_H\right>-0.596)^2+(\left<p_{\bar{H}}\right>-0.374)^2+(1-\left<p_H\right>-\left<p_{\bar{H}}\right>-0.03)^2<\epsilon$, where $\left<p_H\right>$ is the average value of $p_H$ for the population and so on. We set $\epsilon=0.0012$, corresponding to parameters that produce a rating distribution within roughly 2\% of the empirical distribution.

The resulting parameter distributions are shown in Figure~\ref{fig:si-distributional-params}. In particular we see that $\mu^I_u=\mu^I_n$ is highly contained to values in the range $0.2-0.4$,and $gamma$ is constrained to values $>10$. The other parameters are relatively uniformly distributed across the full range of possible values. 

\begin{figure*}
\centering
\includegraphics[width=0.5\textwidth]{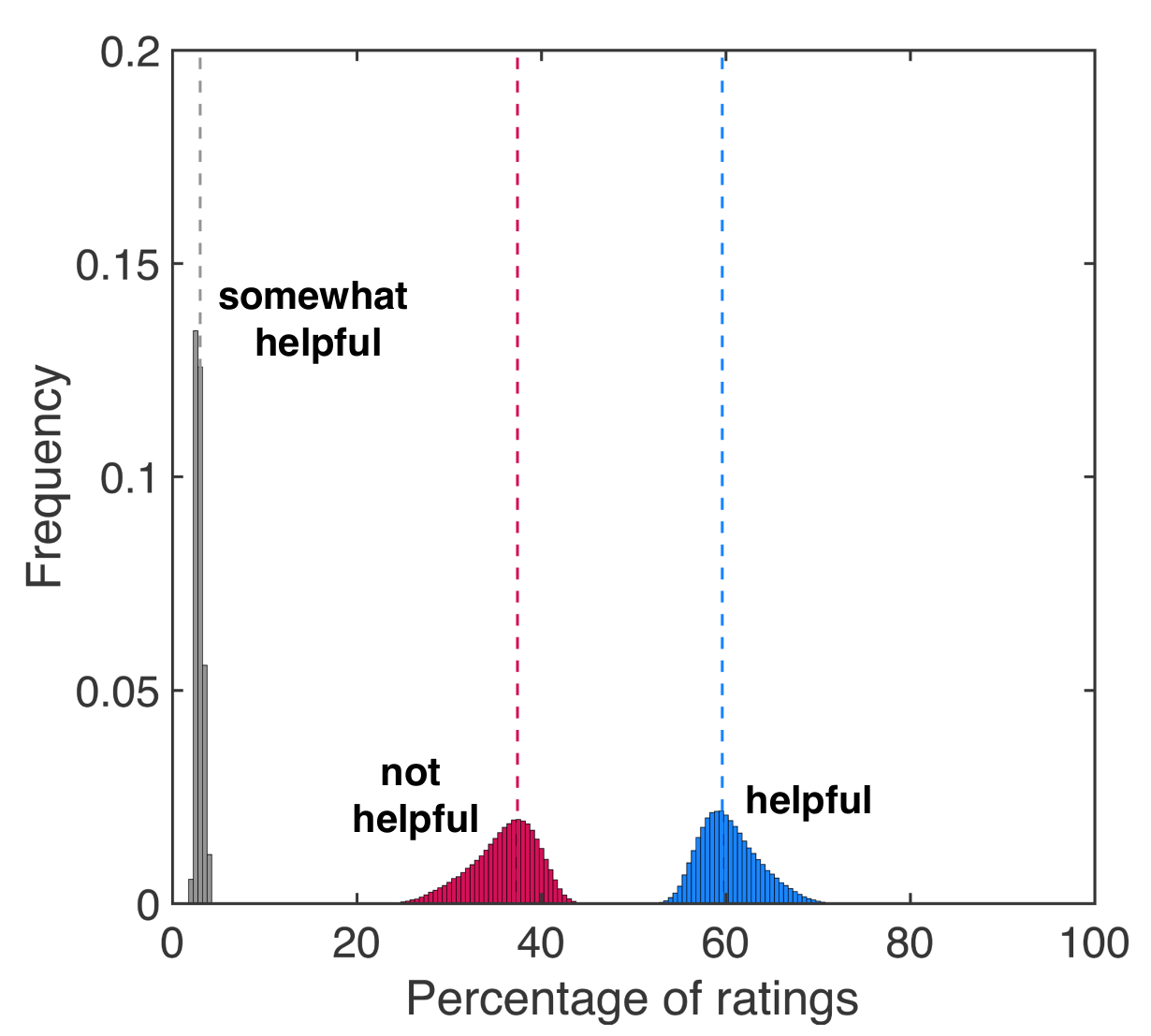}
\caption{\textbf{Ratings distribution for default parameters.} We chose default parameters, $\mu^I_u=\mu^I_n=0.25$, $\gamma=30$, and $\sigma^I_n=0.5$. We then drew sampled $10^7$ parameter sets for the five remaining parameters $\mu^+_u=-\mu^-_u$, $\mu^+_n=-\mu^-_n$, $\sigma^+_u=\sigma^-_u$, $\sigma^+_n=\sigma^-_n$, and $\sigma^I_u$, and calculated the rating proportions for each one. The resulting distributions, as shown in the figure, are closely centered around the empirical rating proportions (dashed lines).}
\label{fig:si-ratings-vs-params}
\end{figure*}

We choose $\mu^I_u=\mu^I_n=0.25$, $\gamma=30$, and $\sigma^I_n=0.5$, which consistently produce rating distributions close to the empirical ones (Figure~\ref{fig:si-ratings-vs-params}). We also choose for our ``baseline condition'' the parameter values $\mu^+_u=\mu^-_u=\mu^+_n=\mu^-_n=0$ and $\sigma^+_u=\sigma^-_u=\sigma^+_n=\sigma^-_n=0.5$ (corresponding to no polarization) and $\sigma^I_0=0.2$, which produces relatively low error rates (Figure ~\ref{fig:variability}b in the main text). In the main text we systematically vary polarization, and rater friendliness. 

\clearpage

\section{Supplementary Results}

In this section we provide additional results to those presented in the main text, including robustness checks and additional experimental manipulations.

\subsection{Waste rate and infiltration rate with bad raters}
\label{sec:si-error-rates}

\begin{figure*}[b!]
\centering
\includegraphics[width=0.6\textwidth]{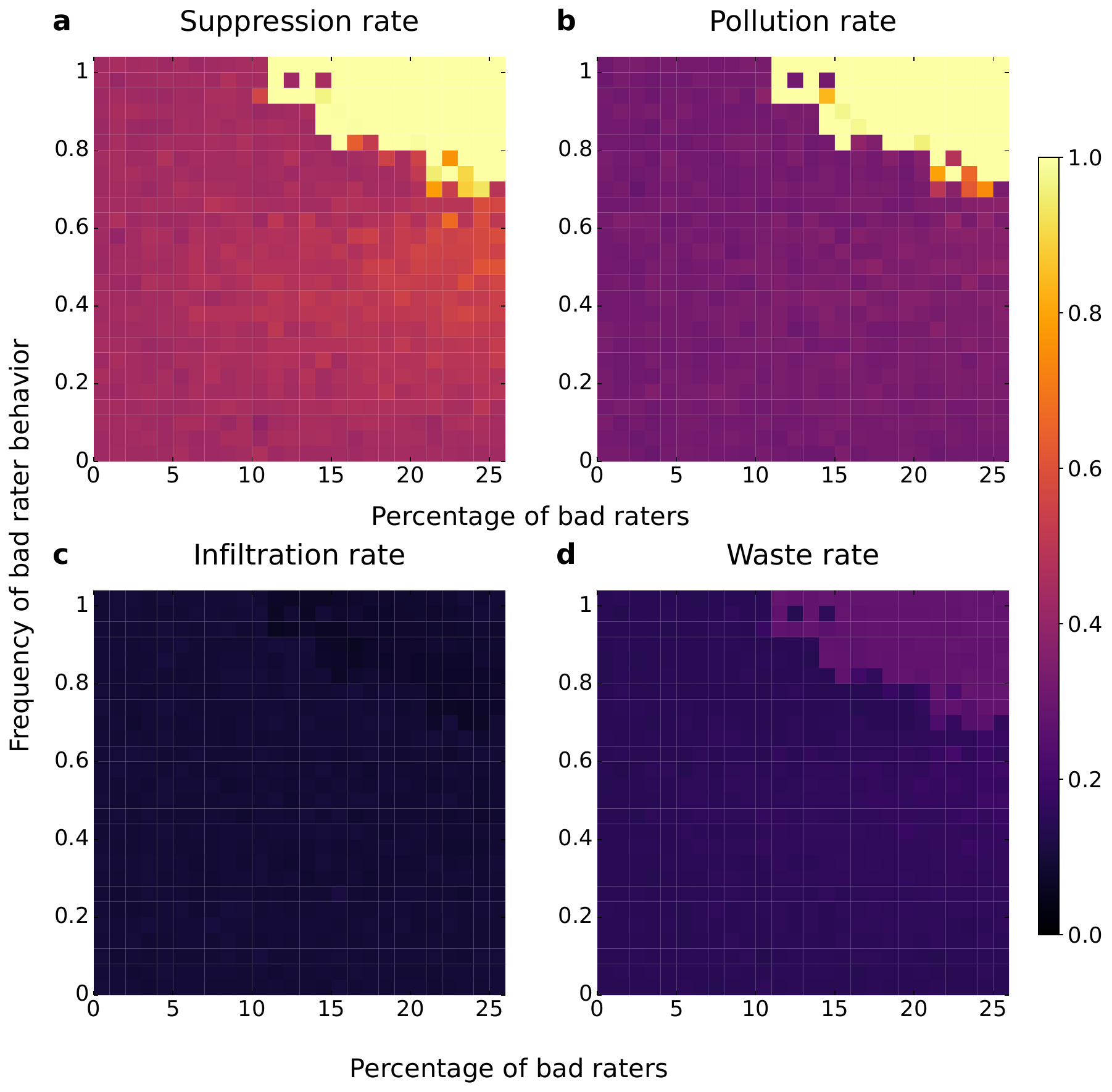}
\caption{\textbf{Indiscriminate bad raters.} This figure shows the same results as Figure~\ref{fig:nonbiased-bad} of the main text (a-b), with additional panels showing (c)~the infiltration rate and (d)~the waste rate.}
\label{fig:si-nonbiased-bad-errors}
\end{figure*}

In the main text we show the impact of indiscriminate bad raters (Figure~\ref{fig:nonbiased-bad}) on the pollution rate and suppression rate of the Community Notes algorithm. Figure~\ref{fig:si-nonbiased-bad-errors} shows, in addition, the waste rate and infiltration rate for the same manipulations. We find that the infiltration rate is not impacted by indiscriminate bad raters, while the waste rate is moderately elevated under the same conditions that produce suppression and infiltration rates close to one.

\begin{figure*}[t]
\centering
\includegraphics[width=0.6\textwidth]{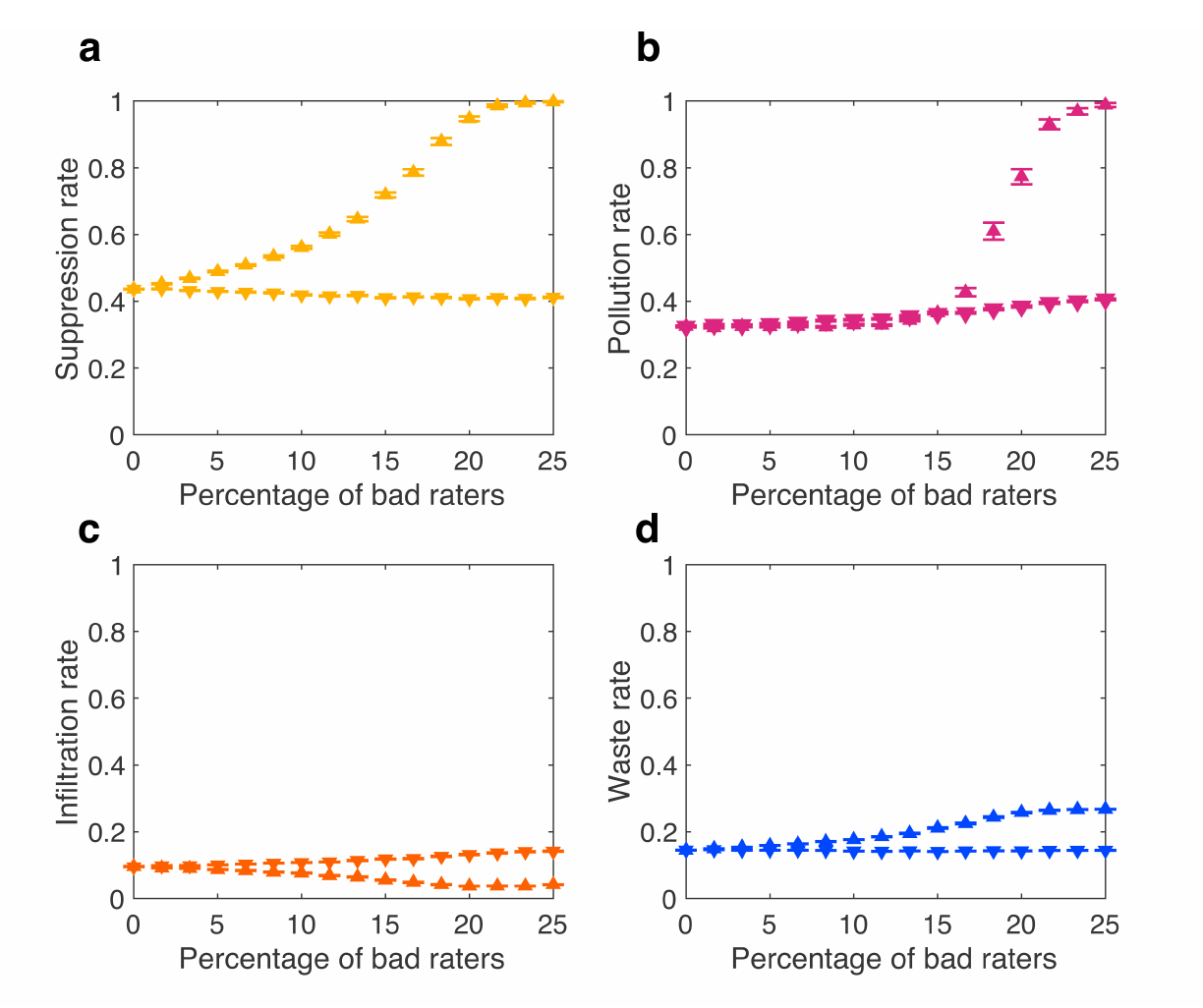}
\caption{\textbf{Coordinated bad raters.} This figure shows the same results as Figure~\ref{fig:biased-bad} of the main text (a-b), with additional panels showing (c)~the infiltration rate and (d)~the waste rate.}
\label{fig:si-biasedbad-errors-x}
\end{figure*}

Similarly, Figure~\ref{fig:si-biasedbad-errors-x} plots the waste and infiltration rates as a function of the proportion of coordinated bad raters who seek to suppress notes with similar bias, under the same conditions as shown in main text (Figure~\ref{fig:biased-bad}. We see that coordinated bad raters have a moderate effect on both waste and infiltration rates.

\begin{figure*}[t]
\centering
\includegraphics[width=\textwidth]{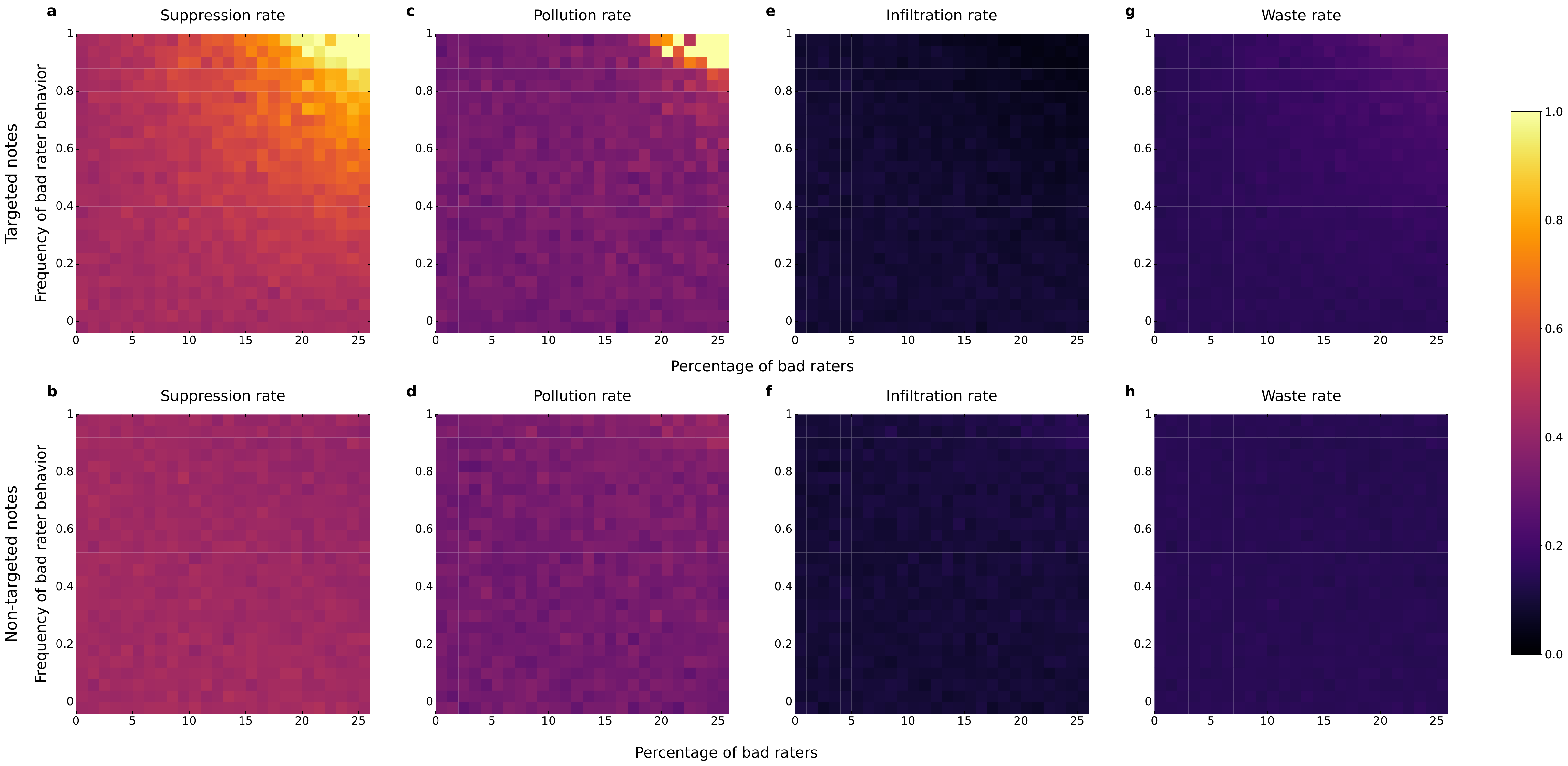}
\caption{\textbf{Coordinated bad raters heat maps.} We systematically varied the percentage of coordinated bad raters and the frequency with which they mis-rate notes they perceive as helpful. (a)~The suppression rate remains relatively low until the percentage of bad raters reaches 20\%, along with a frequency of bad rater behavior around 0.8, after which the suppression rate reaches 100\% for targeted notes, indicating that \emph{all} truly helpful notes go unpublished. (b)~Non-targeted notes have a constant suppression rate. (c)~A similar pattern occurs for the pollution rate, indicating that \emph{none} of the published notes are truly helpful. (d)~Non-targeted notes have a constant pollution rate. (e-f)~The infiltration rate of both targeted (e) and non-targeted (f) notes is largely constant. (g-h)~Similarly, the waste rate of both targeted (g) and non-targeted (h) notes is largely constant.
Results shown are for a single replicate with 20,000 notes and 10,750 raters, for each choice of bad rater percentage and frequency of bad rater behavior. Parameters are as described in Figure~\ref{fig:biased-bad} in the main text, for unpolarized populations ($\rho_n=\rho_u=0$) and with no in-group bias ($E_h=0$).}
\label{fig:si-biasedbad-filter-true}
\end{figure*}

Finally, Figure~\ref{fig:si-biasedbad-filter-true} shows heat maps analogous to those shown in Figures~\ref{fig:nonbiased-bad} and \ref{fig:si-nonbiased-bad-errors}, but for the case of coordinated bad raters, and separating the effects on targeted versus non-targeted notes. 
We observe similar patterns to those reported for indiscriminate bad raters, namely that when both bad behavior rate and the percentage of bad raters reach a threshold, targeted notes experience high suppression rates and pollution rates, while non-targeted notes are largely unaffected. On the other hand, infiltration and waste rates are largely insensitive to coordinated bad raters.

\clearpage

\subsection{Helpfulness filter efficacy with rater and note bias}
\label{sec:si-robustness-distr-effects}

\begin{figure*}[b!]
\centering
\includegraphics[width=0.6\textwidth]{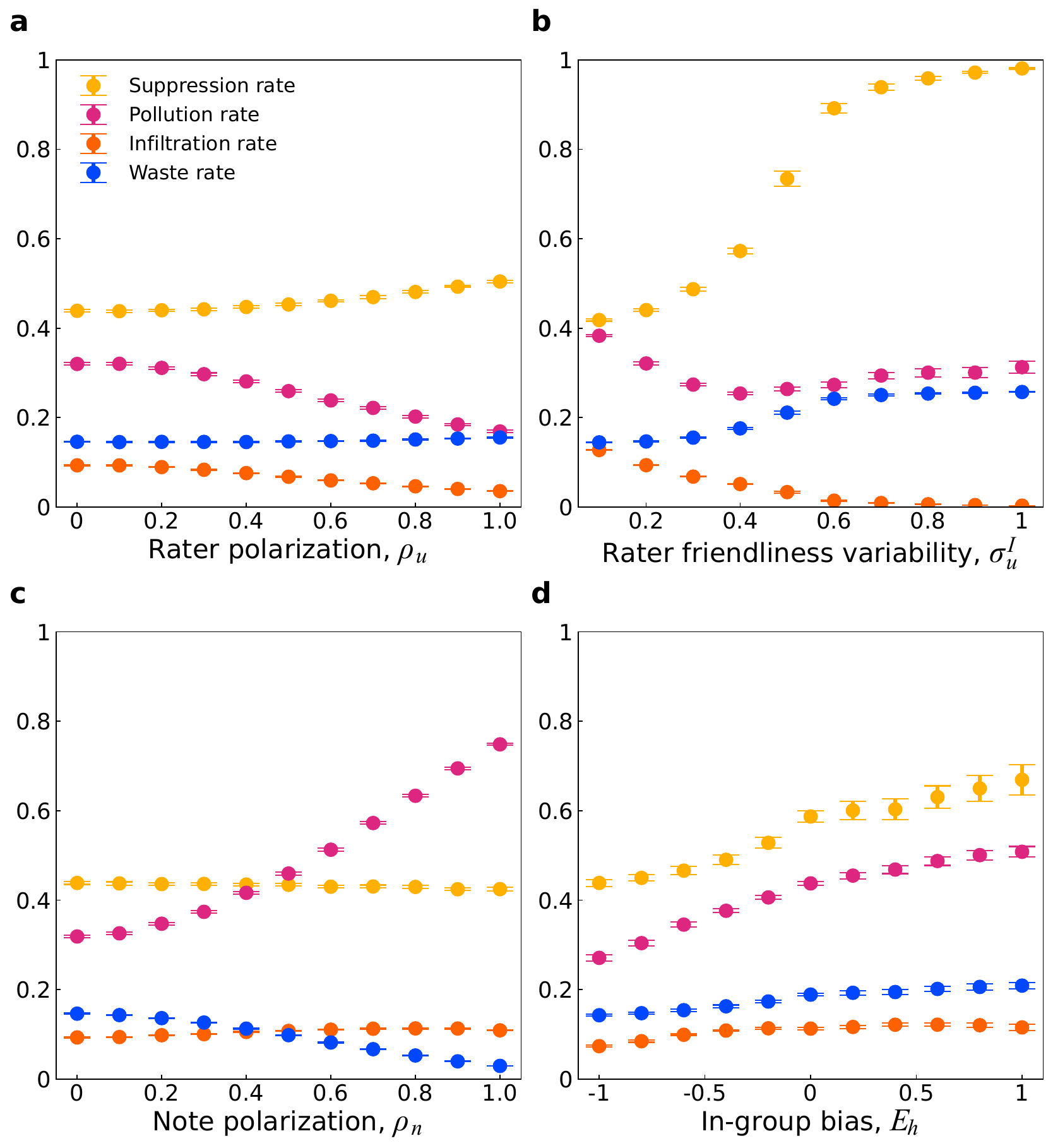}
\caption{\textbf{Varying the distributions of raters and notes without helpfulness filter.} This is the same analysis as shown in Figure~\ref{fig:variability}, with the helpfulness filter turned off.
(a)~Effects of rater polarization $\rho_u$ with unpolarized notes. (b)~Effects of rater friendliness variability, as measured by standard deviation $\sigma^I_u$, in unpolarized populations. (c)~Effects of note polarization $\rho_n$ with unpolarized raters. (d)~Effects of in-group bias $E_h$ when choosing which notes to rate in unpolarized populations. Plots show mean and standard errors across 50 replicate datasets of 20,000 notes and 10,750 raters, matching the real-world distribution of ratings (see \nameref{sec:methods}). In all cases we set $\mu^I_n = \mu^I_u = 0.25$ and $\sigma^I_n = 0.5$, reproducing the empirical frequency of HELPFUL ratings.}
\label{fig:si-fig2}
\end{figure*}

In the main text (Figure~\ref{fig:variability}) we present results on the effects of rater and note polarization, in-group bias, and rater friendliness variability on the rates of errors made by the Community Notes algorithm. Figure~\ref{fig:si-fig2} shows the same analyses with the helpfulness filter turned off. We see that the results are qualitatively the same in both cases.

\clearpage

\subsection{Helpfulness filter efficacy with bad raters}
\label{sec:si-robustness-bad-raters}

\begin{figure*}[b!]
\centering
\includegraphics[width=\textwidth]{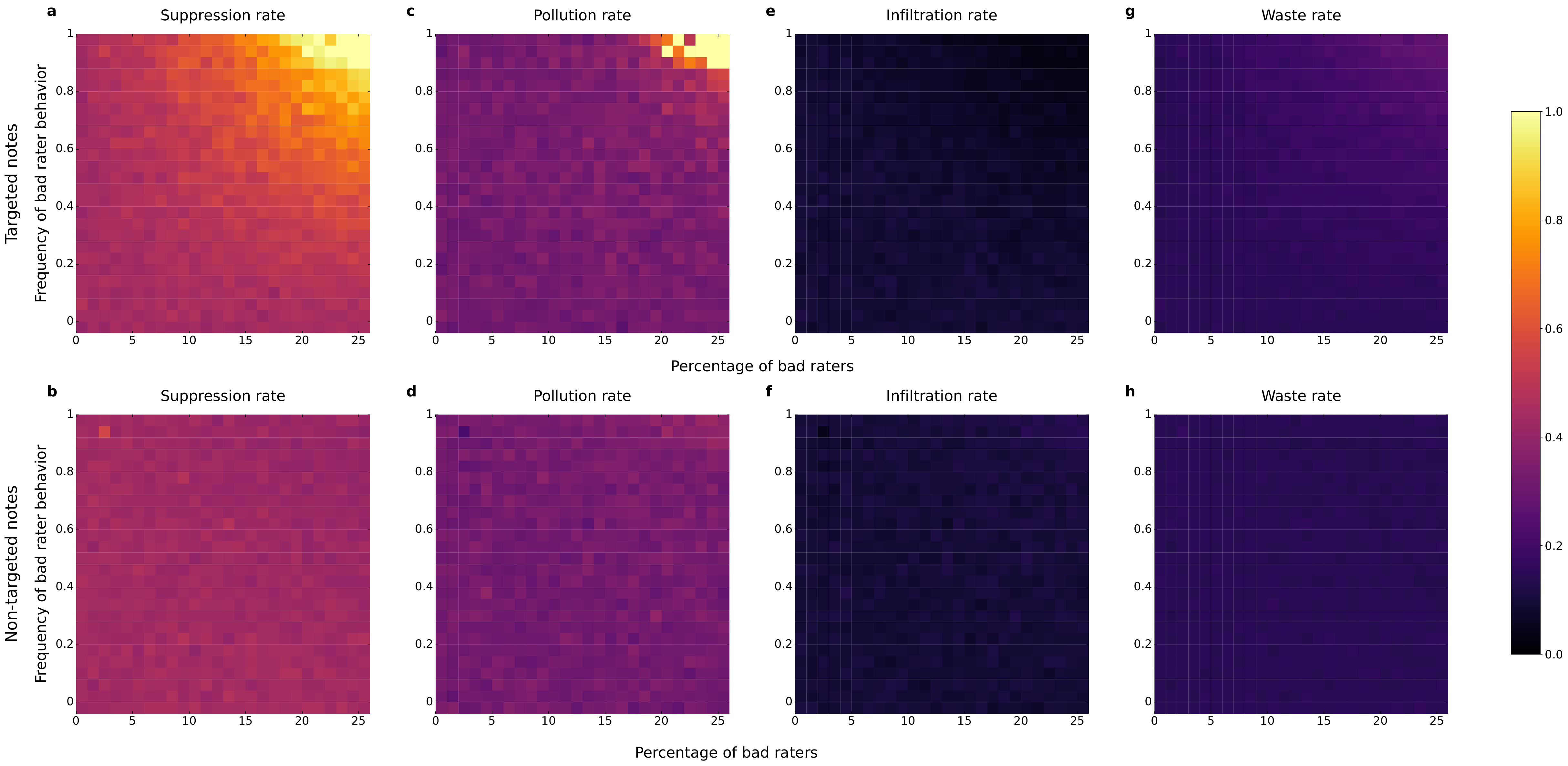}
\caption{\textbf{Coordinated bad raters without helpfulness filter.} This is the same analysis as shown in Figure~\ref{fig:si-biasedbad-filter-true}, with the helpfulness filter turned off.}
\label{fig:si-biasedbad-filter-false}
\end{figure*}

We assessed the impact of the helpfulness filter in the presence of bad raters. Figure~\ref{fig:si-biasedbad-filter-false} shows how error rates are affected by the percentages of coordinated bad raters with the helpfulness filter turned off. This is in contrast to the analysis in Figure~\ref{fig:si-biasedbad-filter-true}, where the helpfulness filter is turned on. We see that the results are largely unaffected by the helpfulness filter. We observe a similar pattern in the case of indiscriminate bad raters.

\clearpage

\subsection{Breakdown of the helpfulness filter}
\label{sec:si-filter-efficacy}

\begin{figure*}[b!]
\centering
\includegraphics[width=0.6\textwidth]{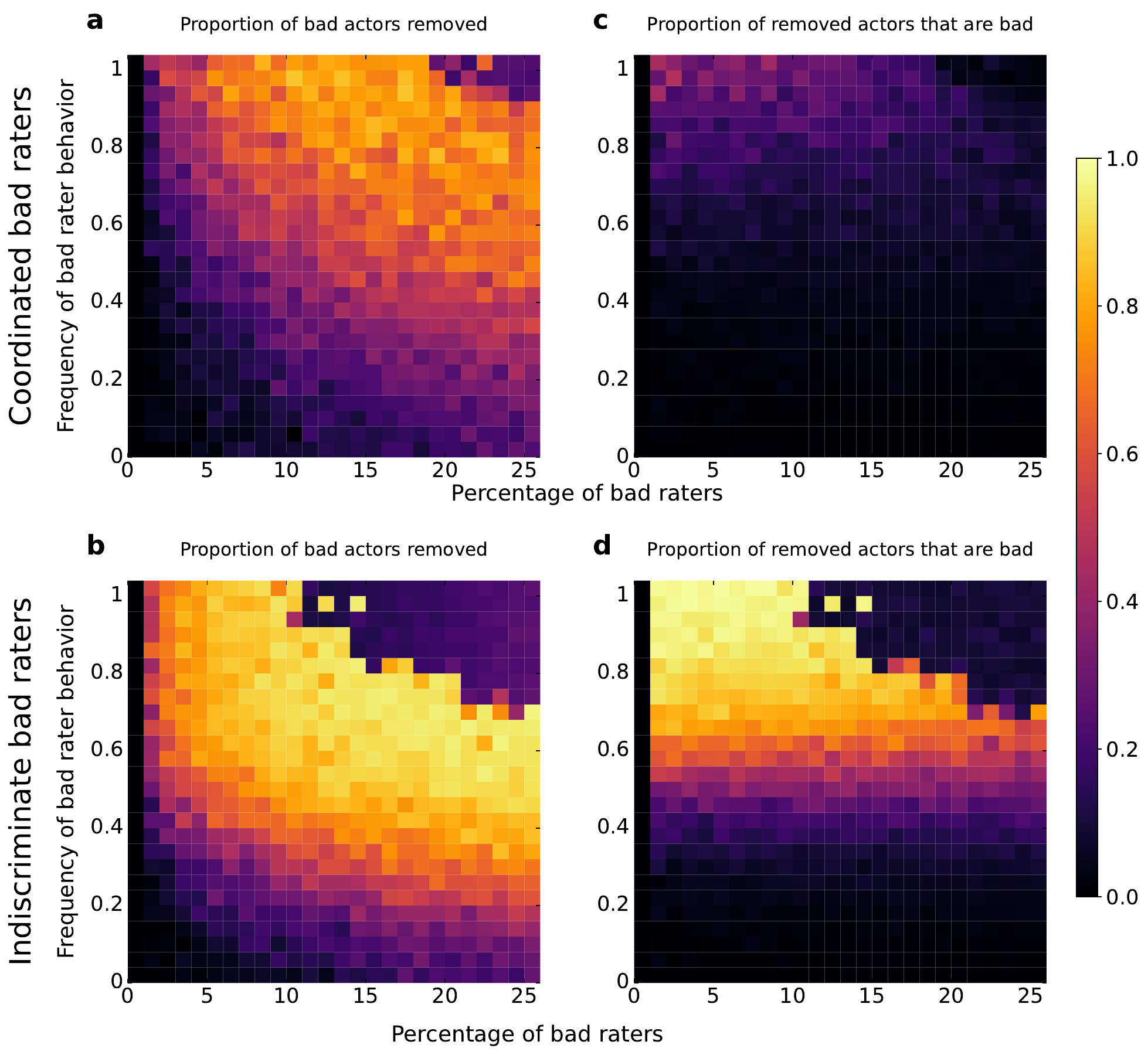}
\caption{\textbf{Helpfulness filter efficacy.} We plot the efficacy of the helpfulness filter as a function of the proportion of bad raters and the frequency of bad rater behavior. We report on two metrics: the proportion of bad raters removed (a, c) and proportion of removed raters that are bad (c, d), in the case of coordinated (top) and indiscriminate (bottom) bad raters. 
}
\label{fig:si-filter-effects}
\end{figure*}

We explored the impact of the helpfulness filter by plotting the rate of removal of both indiscriminate and coordinated bad raters (Figure~\ref{fig:si-filter-effects}). These experiments correspond to those shown in Figure~\ref{fig:nonbiased-bad} of the main text and Figure~\ref{fig:si-biasedbad-filter-true}, respectively. In both cases we find that the proportion of bad raters removed is high (i.e., almost all bad raters are removed) up to a critical threshold, at which point the filter breaks down and bad raters are no longer removed.
In the case of coordinated bad raters, we observe another problem: even when the helpfulness filter is able to remove a good portion of bad raters, most of the removed raters are false positives (honest raters).  

\clearpage

\subsection{Alternate measures of algorithm efficacy}
\label{sec:si-excess-helpfulness-bias}

In the main text we primarily focus on the rates of errors made by the algorithm. Here we explore alternate measures of algorithm efficacy. Results are shown varying the percentage of coordinated bad raters in populations without polarization or in-group bias (i.e., the scenario of Figure~\ref{fig:biased-bad} in the main text). 

\begin{figure*}[b!]
\centering
\includegraphics[width=0.6\textwidth]{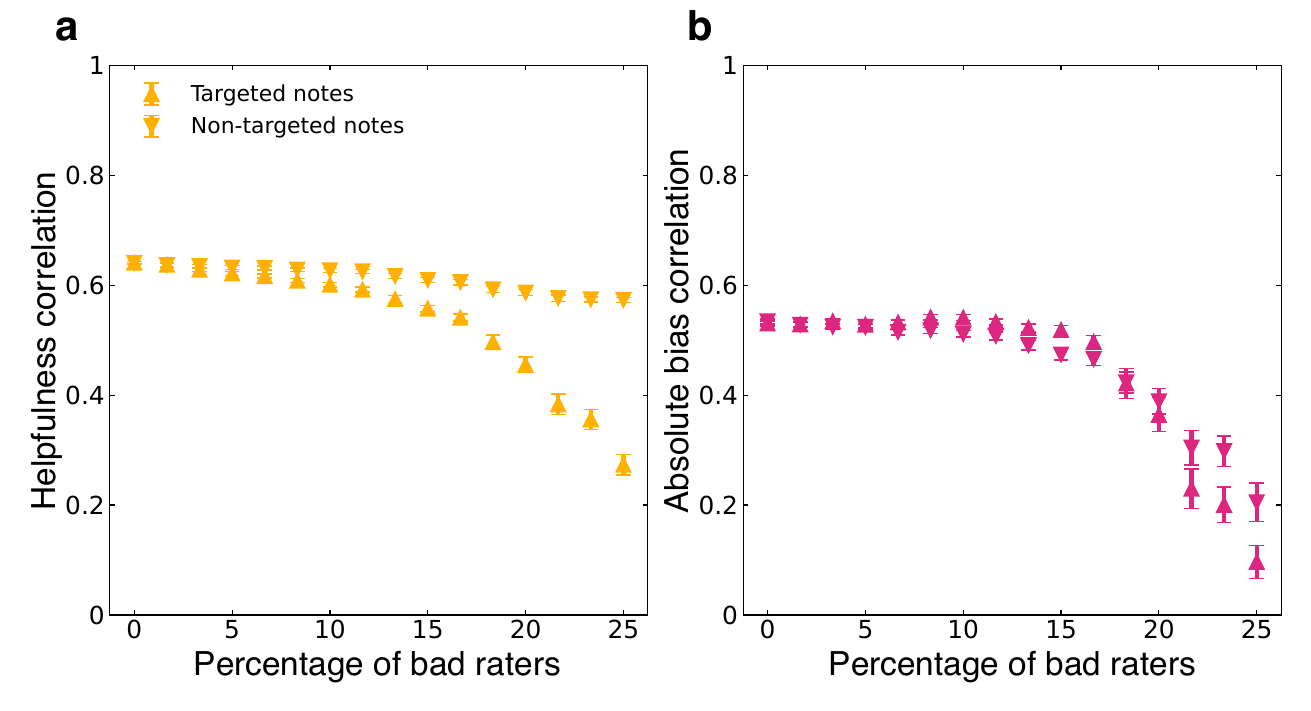}
\caption{\textbf{Correlation between the true and inferred parameters with coordinated bad raters}. Shown are the average Pearson correlation coefficients between (a) the true and inferred helpfulness and (b) the true and inferred absolute bias of notes, in the experiments reported in main text Figure~\ref{fig:biased-bad}.}
\label{fig:si-biasedbad-corr}
\end{figure*}

We first calculate the Pearson correlation between the true and inferred note helpfulness, $i_n$ and $\hat{i}_n$, and for the true and inferred absolute note bias, $|f_{n}|$ and $|\hat{f}_{n}|$. We see in Figure~\ref{fig:si-biasedbad-corr} that the helpfulness correlation declines as the percentage of bad raters increases, moderately for non-targeted notes and sharply for targeted notes, once the percentage of bad raters exceeds a critical threshold --- around 15\% of the population. Above the threshold, the correlation between true and inferred absolute bias also declines sharply, for both targeted and non-targeted notes. These results are consistent with the breakdown of the Community Notes algorithm shown in Figure~\ref{fig:si-biasedbad-errors-x}.

\begin{figure*}
\centering
\includegraphics[width=0.6\textwidth]{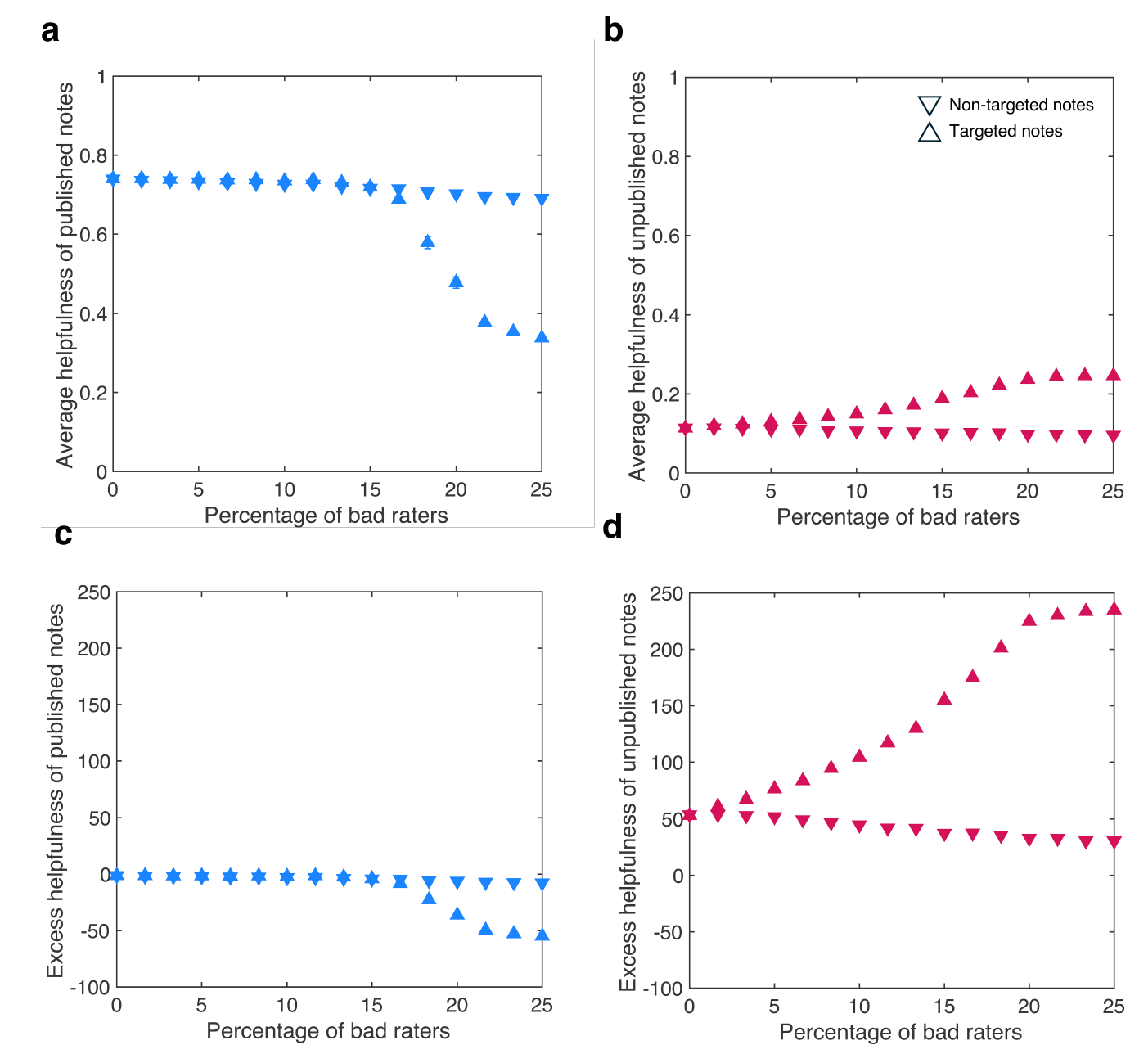}
\caption{\textbf{Helpfulness of targeted and non-targeted notes.} Shown are the average (a-b) and excess (c-d) helpfulness of published (left) and unpublished (right) notes for the experiment reported in Figure~\ref{fig:biased-bad} in the main text.}
\label{fig:si-biasedbad-helpfulness-avg}
\end{figure*}

\begin{figure*}
\centering
\includegraphics[width=0.6\textwidth]{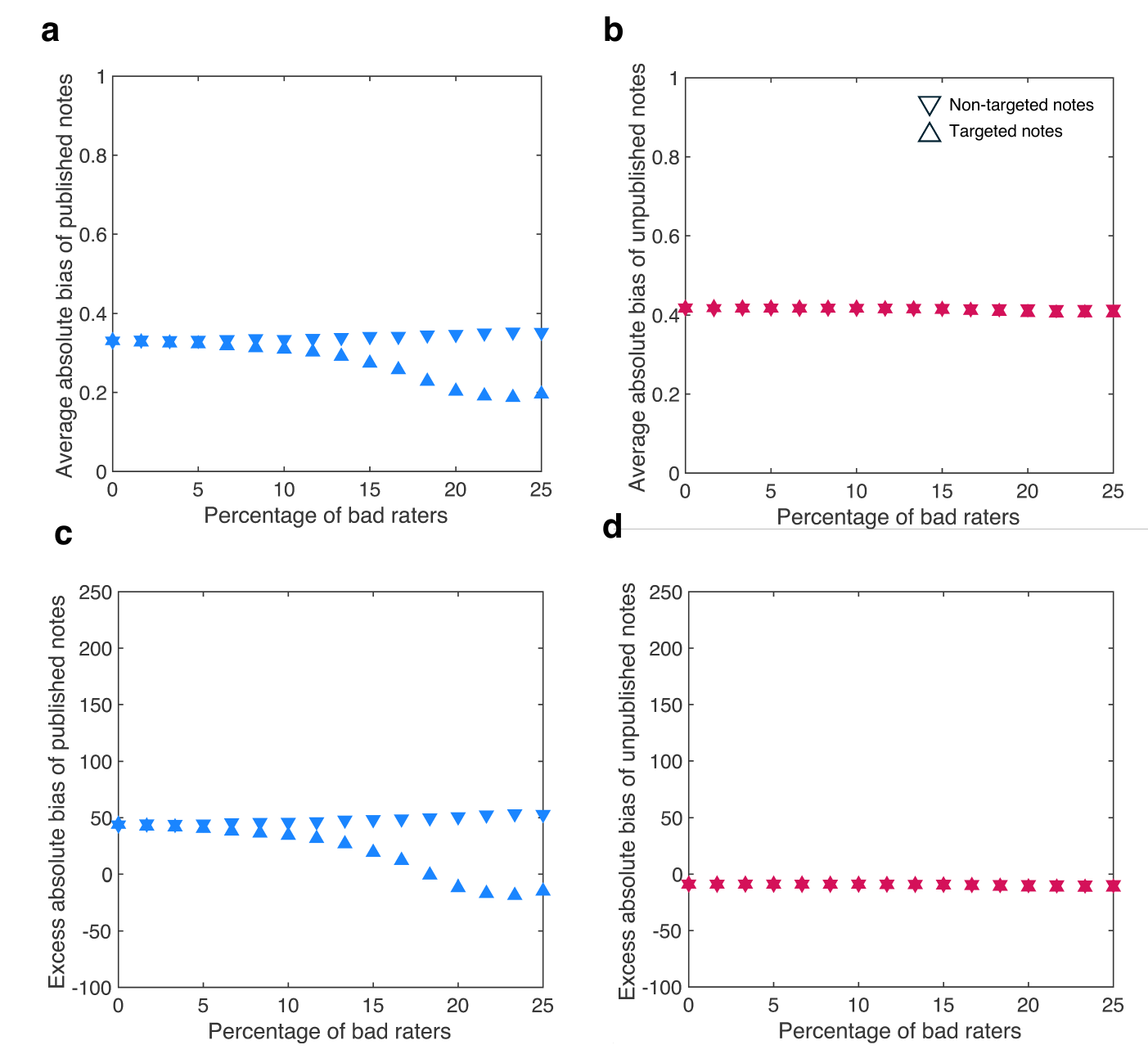}
\caption{\textbf{Absolute bias of targeted and non-targeted notes.} Shown are the average (a-b) and excess (c-d) absolute bias of published (left) and unpublished (right) notes for the experiment reported in Figure~\ref{fig:biased-bad} in the main text.}
\label{fig:si-biasedbad-bias-avg}
\end{figure*}


We also calculated the average true helpfulness of published and unpublished notes, as well as the excess helpfulness of published and unpublished notes (Figure~\ref{fig:si-biasedbad-helpfulness-avg}), for the experiment reported in Figure~\ref{fig:biased-bad} in the main text. We see that targeted notes that are published experience a sharp decline in their average helpfulness and their excess helpfulness, while the helpfulness of non-targeted notes experiences only a very slight decline. In contrast, unpublished targeted notes experience a sharp increase in average and excess helpfulness, while non-targeted notes are largely unaffected. 
These results are consistent with the high suppression rate reported in Figure~\ref{fig:biased-bad}. 

We conducted the same analysis for the absolute bias of targeted and non-targeted notes (Figure~\ref{fig:si-biasedbad-bias-avg}). We observe a decline in the bias of targeted notes that are published, while non-targeted notes and unpublished notes are unaffected. 



\begin{figure*}[b!]
\centering
\includegraphics[width=0.6\textwidth]{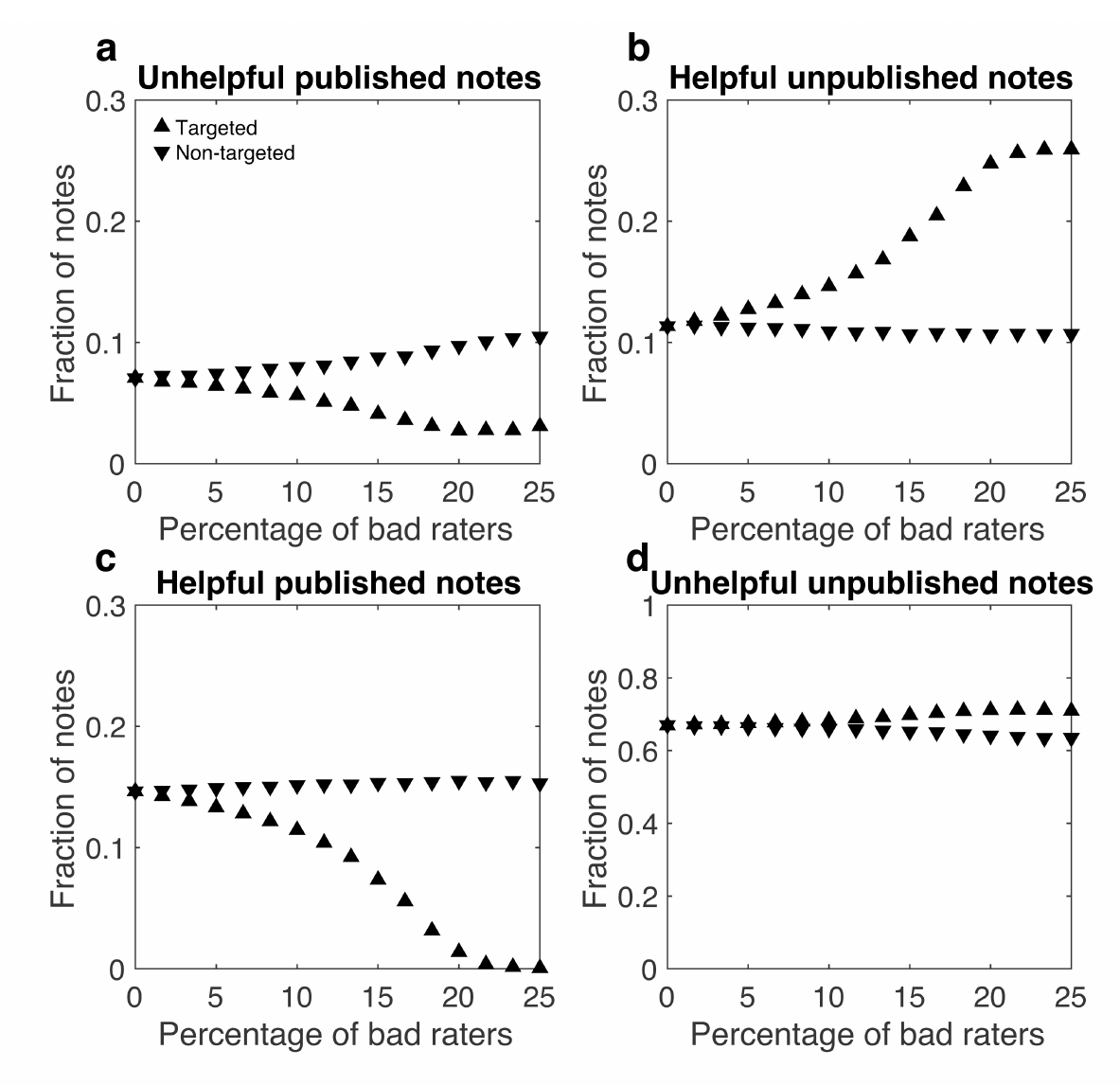}
\caption{\textbf{Effect of coordinated bad raters on different groups of notes.} The plots show the fractions of notes in four groups, based on whether they are helpful and whether they are published, for both targeted and non-targeted notes.}
\label{fig:si-biasedbad-note-fraction}
\end{figure*}

Finally, we calculated the fraction of notes that belong to each of four categories: (i)~unhelpful (i.e., not publishable) but published, (ii)~helpful (i.e., publishable) but unpublished, (iii)~helpful (i.e., publishable) and published, and (iv)~unhelpful (i.e., unpublishable) and unpublished. Figure~\ref{fig:si-biasedbad-note-fraction} plots these fractions as a function of the percentage of coordinated bad raters. We see that among targeted notes, the number of helpful published notes declines to zero as the percentage of bad raters increases, while the number of unhelpful published notes declines but remains positive. In contrast, the number of helpful unpublished notes increases. Among non-targeted notes, there is much less change. 

\clearpage

\subsection{Effects of dataset size}
\label{sec:si-robustness-datasize}

\begin{figure*}[b!]
\centering
\includegraphics[width=0.7\textwidth]{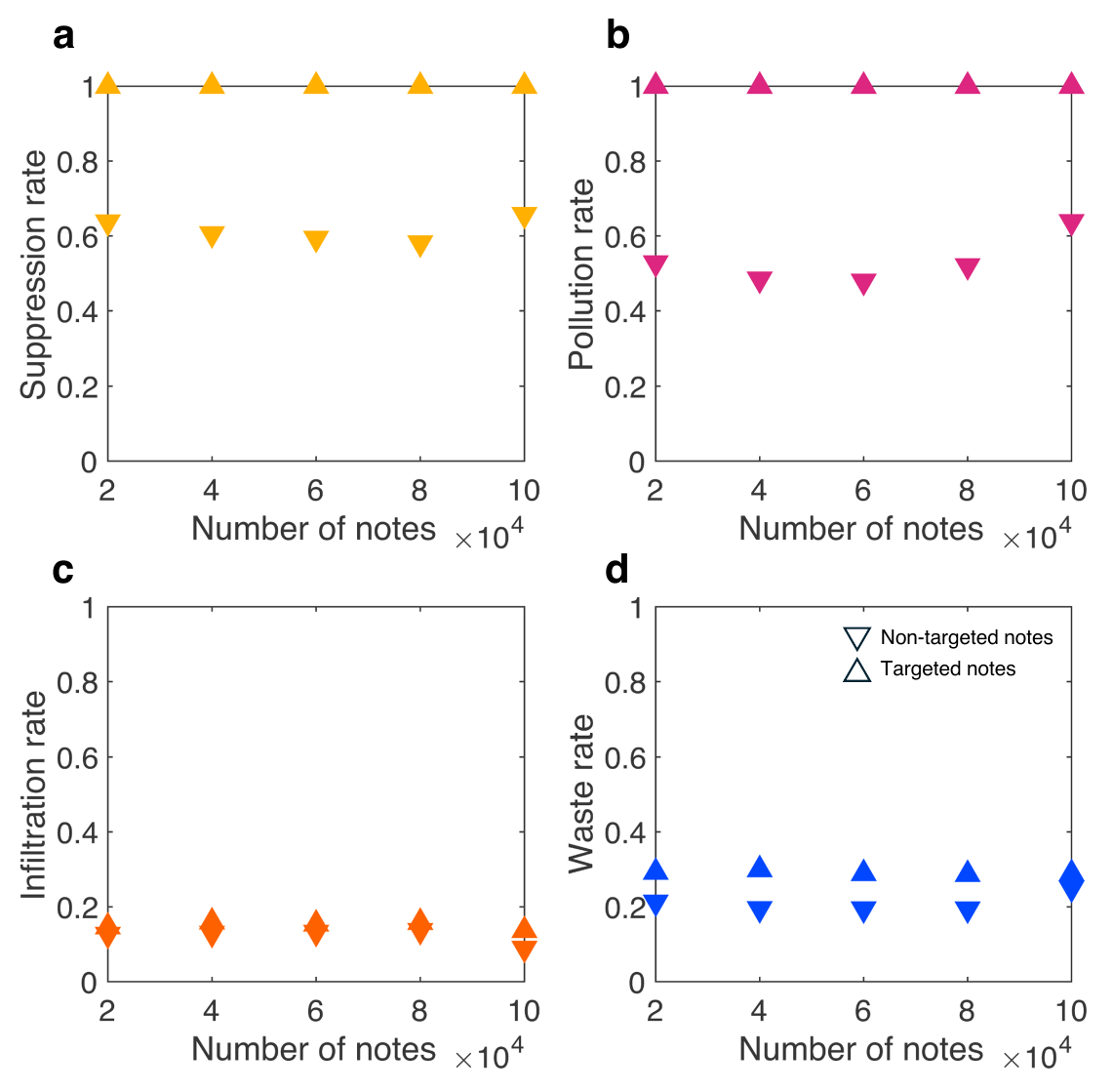}
\caption{\textbf{Dataset size.} Error rates for single replicates with different dataset sizes. Here we assume 25\% bad raters in an experiment with no polarization and the same distribution parameters as used in Figure~\ref{fig:biased-bad}, with the exception that here we set $E_h=1$ for convenience.}
\label{fig:si-size-robustness}
\end{figure*}

We explored the impact of the dataset size on our results by constructing rater-note networks in the same way as described in the \nameref{sec:methods}, for populations of up to 100,000 notes. We ran these simulations for populations with no rater or note polarization, in-group bias $E_h=1$, and 25\% coordinated bad raters. Figure~\ref{fig:si-size-robustness} shows that the error rates are largely insensitive to the dataset size between 20,000 and 100,000 notes. 

\end{document}